\documentclass[%
reprint,
 amsmath,amssymb,
aps,
pra,
notitlepage,
]{revtex4-2}
\usepackage{graphicx}
\usepackage{dcolumn}
\usepackage{bm}
\usepackage{hyperref}
\usepackage{textgreek}
\usepackage{xcolor}
\usepackage[caption=false]{subfig}

\begin{document}
\preprint{APS/123-QED}

\title{Vectorial Light in Fabry-P\'erot Resonators in the Normal Dispersion Regime}

\author{Graeme N. \surname{Campbell$^{1,2}$}}
\email{graeme.campbell.2019@uni.strath.ac.uk}
\author{Lewis \surname{Hill$^{2}$}}
\author{Pascal \surname{Del'Haye$^{2,3}$}}
\author{Gian-Luca \surname{Oppo$^{1}$}}
\affiliation{$^1$SUPA and Department of Physics, University of Strathclyde, Glasgow, G4 0NG, Scotland, UK\\ 
$^2$Max Planck Institute for the Science of Light, 91058 Erlangen, Germany\\ $^3$Department of Physics, Friedrich-Alexander-Universit\"at Erlangen-N\"urnberg, 91058 Erlangen, Germany
}

\begin{abstract}
	The ranges of existence and stability of dark cavity-soliton stationary states in a Fabry-P\'erot resonator with a Kerr nonlinear medium, vectorial polarization components and normal dispersion are determined. The Fabry-P\'erot configuration introduces nonlocal coupling that shifts the cavity detuning by the round trip average power of the intracavity field. When compared with ring resonators, nonlocal coupling leads to strongly detuned dark cavity solitons that exist over a wide range of detunings. We study symmetry breaking between fields of opposite circular polarization characterized by a codimension-2 bifurcation point unique to the regime of normal group velocity dispersion. We show the spontaneous formation of regular dark soliton crystals separated by Turing patterns of alternating polarization via `self crystallization' due to long range interactions. Frequency combs of dark soliton crystals of two orthogonal polarizations in Fabry-P\'erot resonators display three separate components corresponding to the cavity repetition rate, the wavelength of the periodic pattern and the soliton lattice spacing. 
	
	The system also displays the formation of stationary and dynamical vectorial dark-bright solitons. These solutions are different from previous realizations with bichromatic driving in ring resonators, are composed of locked switching fronts and can undergo Hopf bifurcations when scanning the detuning. Interacting oscillating dark-bright solitons display anti-phase dynamics that changes first into quasi-periodic oscillations and then into in-phase dynamics when increasing the cavity length.   
\end{abstract}


\maketitle



\section{Introduction}\label{sec:intro}
The generation of optical frequency combs \cite{PasquaziReview18} is a active area of research due to the wide range of practical applications that span across various fields including telecommunication \cite{pfeifle2014coherent,pfeifle2015optimally}, spectroscopy \cite{suh2016microresonator,dutt2018chip}, astronomy \cite{Suh2019,Obrzud2019}, LIDAR \cite{Kuse2019} and quantum technologies \cite{reimer2016generation}. Temporal cavity solitons (TCS) \cite{coen2016temporal,Oppo2024} have attracted significant interest for their capability to produce broadband optical frequency combs. TCS are a special class of cavity solitons that originate in dissipative optical resonators under the action of external driving, diffraction \cite{scroggie1994,firth1996} and/or group velocity dispersion \cite{Oppo2024}. Although ring resonator geometries are now regularly used for the generation of optical frequency combs via TCS, there is an increasing interest in Fabry-Pérot (FP) configurations with recent theoretical \cite{cole2018theory,campbell2023dark,hill2024symmetry} and experimental \cite{obrzud2017temporal,wildi2023soliton} studies addressing the generation of bright and dark TCSs. In terms of applications, FP configurations, especially with chirped mirrors, offer superior ways to control dispersion in integrated microresonators, thus enabling broader frequency comb bandwidths \cite{wildi2023soliton}.

\begin{figure}[h!]
    \centering
    \includegraphics[width=0.85\linewidth]{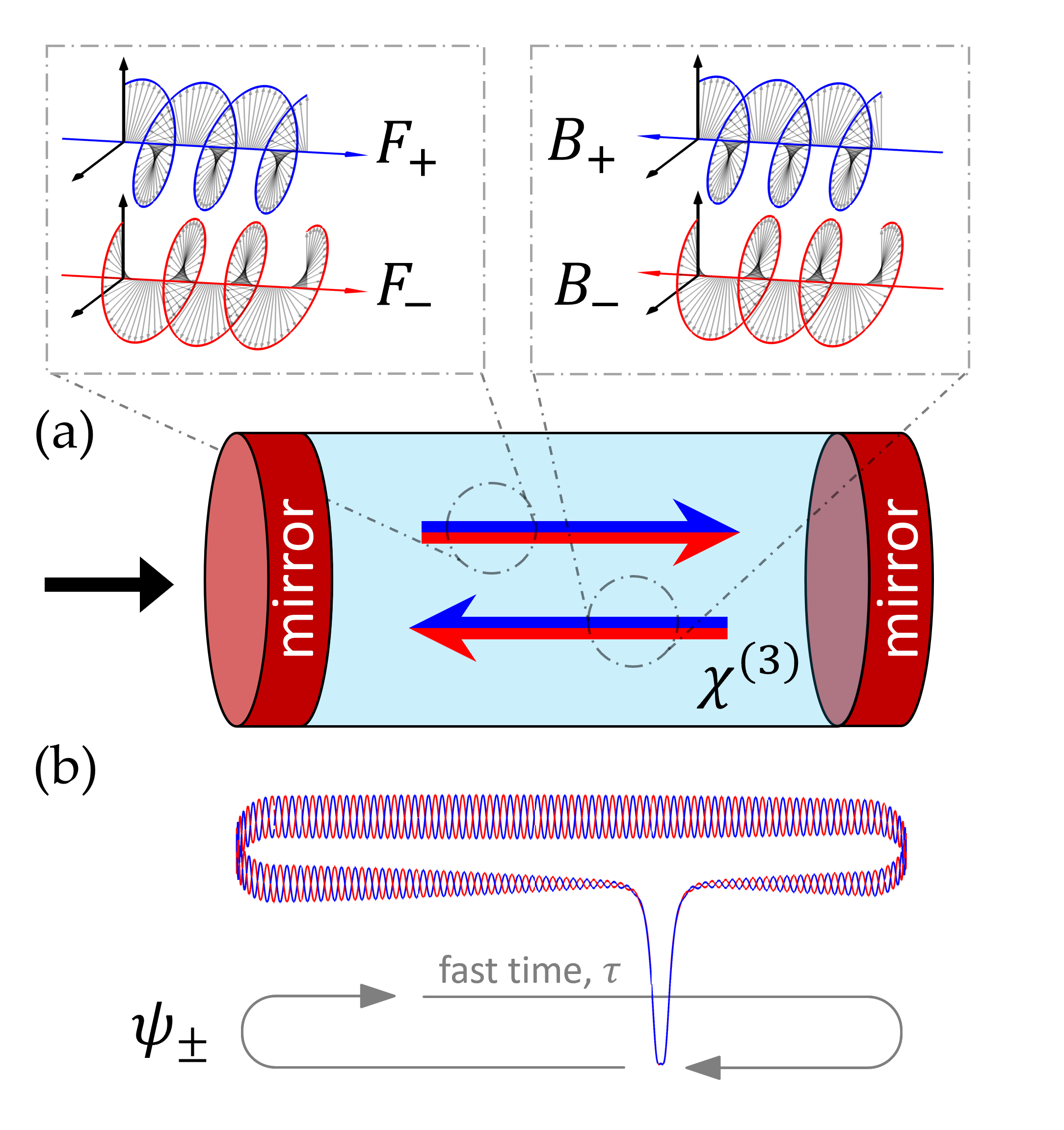}
    \caption[Fabry-P\'erot configuration with two orthogonal polarizations]{(a) A Fabry-P\'erot resonator composed of two highly reflective mirror filled with a Kerr, $\chi^{(3)}$, nonlinear medium. A linearly polarized input field is coupled into the resonator, where the forward, $F_\pm$, and backward, $B_\pm$, counterpropagating fields are be resolved into orthogonal polarization components `$+$', `$-$'. (b) An example vectorial dark soliton solution is presented in terms of the auxiliary fields $\psi_\pm$.}
    \label{fig:setup_FP2pol}
\end{figure}

Here we consider a high finesse FP resonator filled with a Kerr medium, see Fig. \ref{fig:setup_FP2pol}, in the normal dispersion regime. A linearly polarized driving laser is coupled to one of the cavity mirrors, such that the intracavity fields may be resolved into counter-propagating components of opposite circular polarization \cite{hill2023multi}. This allows for additional properties of the TCS, as those seen for the single field FP \cite{cole2018theory,campbell2023dark}, due to the possibility of spontaneous symmetry breaking (SSB) between polarizations \cite{Xu21,Faticon24}. We investigate the polarization properties of the FP TCSs in the normal dispersion regime and their effects on the formation of frequency combs. In particular, we present a mechanism of self crystallization in which an initial random distribution of dark solitons ends spontaneously to form a regular soliton crystal \cite{campbell2024frequency}. This is achieved without the use of any external modulation of the field \cite{lu2021synthesized}, but via SSB and long range interactions typical of photonics configurations with counter propagation. 

We begin in Section \ref{sec:FPSSB_derivation} by deriving our model for the FP resonator with two orthogonal polarization components, as was first derived in \cite{hill2024symmetry}. Following the work of Pitois el al. \cite{pitois2001nonlinear} we describe the interaction of counter-propagating vectorial fields in a Kerr nonlinear medium. Then, by taking inspiration form Cole et al. \cite{cole2018theory}, we combine two forward and two backward propagating field equations to obtain a pair of mean field integro-partial differential equations describing fields of counter-rotating circular polarization over the round trip of the cavity. In section \ref{sec:FPSSB_HSS} we discuss the homogeneous stationary solutions (HSS) of this system. In particular, we present a codimension-2 \cite{CodimBif} spontaneous symmetry bifurcation of the high power bistable HSS. This bifurcation represents the collision of a forward supercritical pitchfork bifurcation (found in the normal dispersion regime) which results in the formation of a Turing pattern, and a reverse supercritical pitchfork bifurcation resulting in symmetry broken HSS. 
This bifurcation structure results in a multitude of different symmetry broken vectorial dark solitons (VDS) stationary states as shown in Section \ref{sec:FPSSB_VDS}. Of particular interest is the Turing instability resulting in the formation of patterns of alternating polarizations on the background plateaus from which the VDS hangs. This SSB phenomenon is shown to lead to long range interactions between adjacent VDSs in the ring resonator model. These long range interactions result in the spontaneous self-organisation of VDSs to form regular soliton crystals (RSC) form an initial random distribution \cite{campbell2024frequency}. In Section \ref{sec:FPSSB_selfcry}, we generalise this `self-crystallization' phenomenon to the FP resonator and discuss the differences between these two systems. Finally, we characterize the formation of dark-bright vectorial solitons in Section \ref{sec:FPSSB_VDBS}. Such solutions are shown to undergo a Hopf bifurcation, resulting in non trivial breathing dynamics.

\section{Modelling the Fabry-P\'erot resonator with two polarization components}\label{sec:FPSSB_derivation}
As first derived in \cite{hill2024symmetry}, the system shown in Fig. \ref{fig:setup_FP2pol} is well described by two coupled integro-partial differential equations
\begin{eqnarray}
    \partial_t \psi_\pm &=& S - (1 + i\theta)\psi_\pm - i\partial^2_\tau\psi_\pm + i\frac{4}{3}\langle \psi_\pm\psi_\mp^*\rangle\psi_\mp\label{eq:FP}\\
    &+& i\frac{2}{3}\{ |\psi_\pm|^2 +  2\langle |\psi_\pm|^2\rangle + 2 |\psi_\mp|^2 + 2\langle |\psi_\mp|^2\rangle\}\psi_\pm\nonumber
\end{eqnarray}
describing the evolution of two fields $\psi_\pm(\tau,t)$ of slowly varying amplitudes and orthogonal polarizations over the extended domain $0\leq\tau\leq \tau_\text{R}$ with periodic boundary conditions, where $\tau_\text{R}$ is the round trip time of the resonator, $S$ is the amplitude of the input field, which is considered to be real and positive, and $\theta$ is the detuning of the input field to the nearest cavity resonance \cite{hill2024symmetry}. Here, $t$ is the `slow time' temporal variable describing the evolution over many round trips of the cavity, while $\tau$ is the `fast time' longitudinal variable describing the evolution over a single round trip of the FP cavity in the normal dispersion case. 

Analogously to what described in \cite{cole2018theory,campbell2023dark,hill2024symmetry}, the fields $\psi_\pm$ of Eqs. (\ref{eq:FP}) are related to the slowly varying field envelopes of the forward, $F_\pm(\tau,t)$, and backward, $B_\pm(\tau,t)$, counter-propagating intracavity fields with orthogonal polarizations through the modal expansions
\begin{eqnarray}
    F_\pm(\tau,t) = \sum_{\mu=-\infty}^{+\infty}f^\pm_\mu(t)e^{-i\alpha_\mu(t-\tau)},\\
    B_\pm(\tau,t) = \sum_{\mu=-\infty}^{+\infty}f^\pm_\mu(t)e^{-i\alpha_\mu(t+\tau)}
\end{eqnarray}
over the domain $0\leq \tau\leq \tau_\text{R}/2$, where the modal coefficients $f^\pm_\mu$ of mode number $\mu$ are defined as
\begin{eqnarray}
    f^\pm_\mu(t) = \frac{1}{\tau_\text{R}}\int_{-\tau_\text{R}/2}^{\tau_\text{R}/2}d\tau e^{-i\alpha_\mu\tau}\psi_\pm(\tau,t)
\end{eqnarray}
with fast time wavenumber $\alpha_\mu = 2\pi\mu/\tau_\text{R}$.

The terms of Eqs. (\ref{eq:FP}) enclosed in angled brackets represent integrals in the fast time variable over a single round trip of the cavity
\begin{eqnarray}
    \langle |\psi_\pm|^2\rangle &=& \frac{1}{\tau_\text{R}}\int_{-\tau_\text{R}/2}^{\tau_\text{R}/2} |\psi_\pm|^2 d\tau, \label{eq:inttermreal}\\
    \langle \psi_\pm\psi_\mp^*\rangle &=& \frac{1}{\tau_\text{R}}\int_{-\tau_\text{R}/2}^{\tau_\text{R}/2} \psi_\pm\psi_\mp^* d\tau.\label{eq:inttermcomp}
\end{eqnarray}
Integral terms arise due to counter-propagation of intracavity fields present in the FP configuration and are the result of rapid phase dynamics of cross-coupling terms between the fields such that they see each other through an average over the round trip \cite{cole2018theory,Kondratiev20,Skryabin:20}. The integral of Eq. (\ref{eq:inttermreal}) corresponds to the average intracavity power of the field over a round trip of the cavity. Nonlocal coupling of this kind is also present in bidirectionally pumped ring resonators with only two coupled equations, one for each counter-propagating field \cite{Kondratiev20,Skryabin:20,campbell2022counterpropagating}, as well as the single field FP configurations \cite{cole2018theory,campbell2023dark}. Of particular importance here is the term given in Eq. (\ref{eq:inttermcomp}) which is not found in the aforementioned systems. This term corresponds to the energy exchange between circular components of the fields \cite{pitois2001nonlinear}. The presence of this term introduces a phase sensitive nonreciprocal nonlocal coupling between the two field components. As previously shown in the anomalous dispersion regime, the nonreciprocity of this term imposes a `global' conformity of spontaneous symmetry breaking vectorial bright solitons \cite{hill2024symmetry}.

Eqs. (\ref{eq:FP}) of the FP resonator are invariant under the exchange of the indices $+$ and $-$, representing a fundamental symmetry of the system. Setting $\psi_+ = \psi_- = \psi$, the coupled Eqs. (\ref{eq:FP}) reduce to the single equation
\begin{align}\label{eq:symFPtwopol}
    \partial_t\psi = S - (1 + i\theta)\psi + 2i(|\psi|^2 + 2\langle|\psi|^2\rangle)\psi - \partial_\tau^2 \psi.
\end{align}
By performing the renormalization $\psi\rightarrow\psi/\sqrt{2}$, $S\rightarrow S/\sqrt{2}$ we obtain the equation of the FP resonator with a single intracavity field \cite{cole2018theory,campbell2023dark}. As a consequence, the symmetric stationary solutions of Eqs. (\ref{eq:FP}) are also stationary solutions Eq. (\ref{eq:symFPtwopol}). This similarity does not necessarily extend to the stability of such solutions, with the possibility of spontaneous symmetry breaking of vectorial dark solitons in the system investigated here.

\section{Homogeneous stationary states}\label{sec:FPSSB_HSS}
We first consider the HSSs of Eqs. (\ref{eq:FP}) by setting all derivatives to zero, $\partial_t\psi_\pm = 0$, $\partial_\tau^2\psi_\pm = 0$. The HSSs can be determined by solving the coupled equations 
\begin{align}
    S^2 = 4H_\pm^3 - 4(\theta - 4H_\mp)H_\pm^2 + ((\theta - 4H_\mp)^2 +1 )H_\pm,\label{eq:PFtwopolHSS}
\end{align}
where the real and imaginary parts are
\begin{align}
    \begin{pmatrix}
        U_{0,\pm}\\
        V_{0,\pm}
    \end{pmatrix}
    = 
    \begin{pmatrix}
        S / R \\
        S (2H_\pm + 4H_\mp - \theta)/ R        
    \end{pmatrix}
\end{align}
where $R=[1+(2H_\pm + 4H_\mp - \theta)^{2}]$.
This result is obtained by recognizing that the integral terms can be trivially evaluated as $\langle |\psi_\pm|^2\rangle = |\psi_\pm|^2, \langle \psi_\pm\psi_\mp^*\rangle = \psi_\pm\psi_\mp^*$ for solutions with a flat profile. Eqs. (\ref{eq:PFtwopolHSS}) admit solutions that are either linearly polarized (symmetric $H_+ = H_-$) or not (broken symmetry $H_+ \neq H_-$). These solutions allow for the SSB of light in the FP resonator, which has been experimentally demonstrated in FP configurations when neglecting dispersion in for example \cite{moroney2022Kerr}. The set of Eqs. (\ref{eq:PFtwopolHSS}), under the renormalization $H_\pm \rightarrow H_\pm/2, S^2 \rightarrow S^2/2$, are mathematically equivalent to those for a ring resonator with two orthogonal polarization components \cite{Geddes1994polarizationpatterns,garbin2020asymmetric,hill2020effects}, as well as two counter-propagating intracavity fields \cite{woodley2018universal,campbell2022counterpropagating,hill2020effects}. This analogy does not necessarily extend to the stability of these stationary solutions. As such, in the followings sections we investigate the different aspects of the SSB bifurcation structure of the HSS, with particular focus on the formation of Turing structures that are phenomenologically similar to those described for ring resonators \cite{campbell2024frequency}.

\subsection{Linear stability of the homogeneous stationary states}
To investigate the stability of the HSSs of this system, we perform a modal expansion of Eqs. (\ref{eq:FP}) and introduce a linear perturbation, as outlined in Appendix \ref{app:stabHSS}. The homogeneous stationary solutions of Eqs. (\ref{eq:FP}) correspond to the modal coefficients 
\begin{equation}
    f_{\mu,s}^\pm = \psi_s^\pm\delta_{\mu,0}
\end{equation}
where $\delta_{\mu,0}$ is the Kronecker delta. We introduce a small perturbation of the form
\begin{equation}
    f_\mu^\pm = \psi_s^\pm \delta_{\mu,0} + \delta f_\mu^\pm,
\end{equation}
and obtain the set of eigenvalues
\begin{align}
    \lambda(\alpha_\mu) = -1 \pm \frac{\sqrt{-A_+B_+ - A_-B_- - \frac{1}{6}(1-\delta_{\mu,0})C\pm Q}}{\sqrt{2}}\label{eq:FPtwopolEigenvalues},
\end{align}
where
\begin{align}
    Q = \bigg[&(A_+B_+ - A_-B_-)^2 + A_+A_-C\\
    &+ (1-\delta_{\mu,0})\frac{C}{9}(3A_+B_+ + B_+B_- + 3A_-B_-)\bigg]^{1/2}\nonumber,
\end{align}
as we first reported in \cite{hill2024symmetry}, and 
\begin{align}
    A_\pm &= -\theta + \alpha^2_\mu + 2\psi_\pm^2 + \frac{4}{3}(2+\delta_{\mu,0})\psi_\mp^2,\\
    B_\pm &= -\theta + \alpha^2_\mu + \frac{2}{3}(5+4\delta_{\mu,0})\psi_\pm^2 + \frac{4}{3}(2+\delta_{\mu,0})\psi_\mp^2,\\
    C &= 64(1+3\delta_{\mu,0})\psi_+^2\psi_-^2.
\end{align}
Due to the presence of the Kronecker delta in Eq. (\ref{eq:FPtwopolEigenvalues}), we must consider the cases of $\mu=0$ and $\mu\neq0$ separately. These two outcomes refer to the scenario in which the perturbation in slow time has a flat profile, $\mu = 0$, or exhibits a sinusoidal fast time component, $\mu \neq 0$, and are made notably distinct from each other due to the counterpropagation of the fields. This is due to the fact that perturbations $\delta f_\mu^\pm$ do not survive the averaging under the integral terms (originating from counterpropagation) should it contain a fast time component. When $\mu = 0$, Eq. (\ref{eq:FPtwopolEigenvalues}) reduces to eigenvalues which, under the appropriate renormalization, are mathematically identical to those seen in the absence of the fast time variable for the ring resonator system of two counterpropagating intracavity fields (see 
\cite{woodley2018universal,campbell2022counterpropagating,hill2020effects}), or two fields of orthogonal polarizations (see \cite{Geddes1994polarizationpatterns,garbin2020asymmetric,hill2020effects}). These cases are discussed in detail for the FP in \cite{moroney2022Kerr} when neglecting dispersion. When $\mu \neq 0$, the eigenvalues indicate the growth/decay of slow time perturbations that are sinusoidal in the fast time variable with wavenumber $\alpha_\mu$ thus indicating the possible formation of Turing patterns.
\begin{figure}
    \includegraphics[width=1\linewidth]{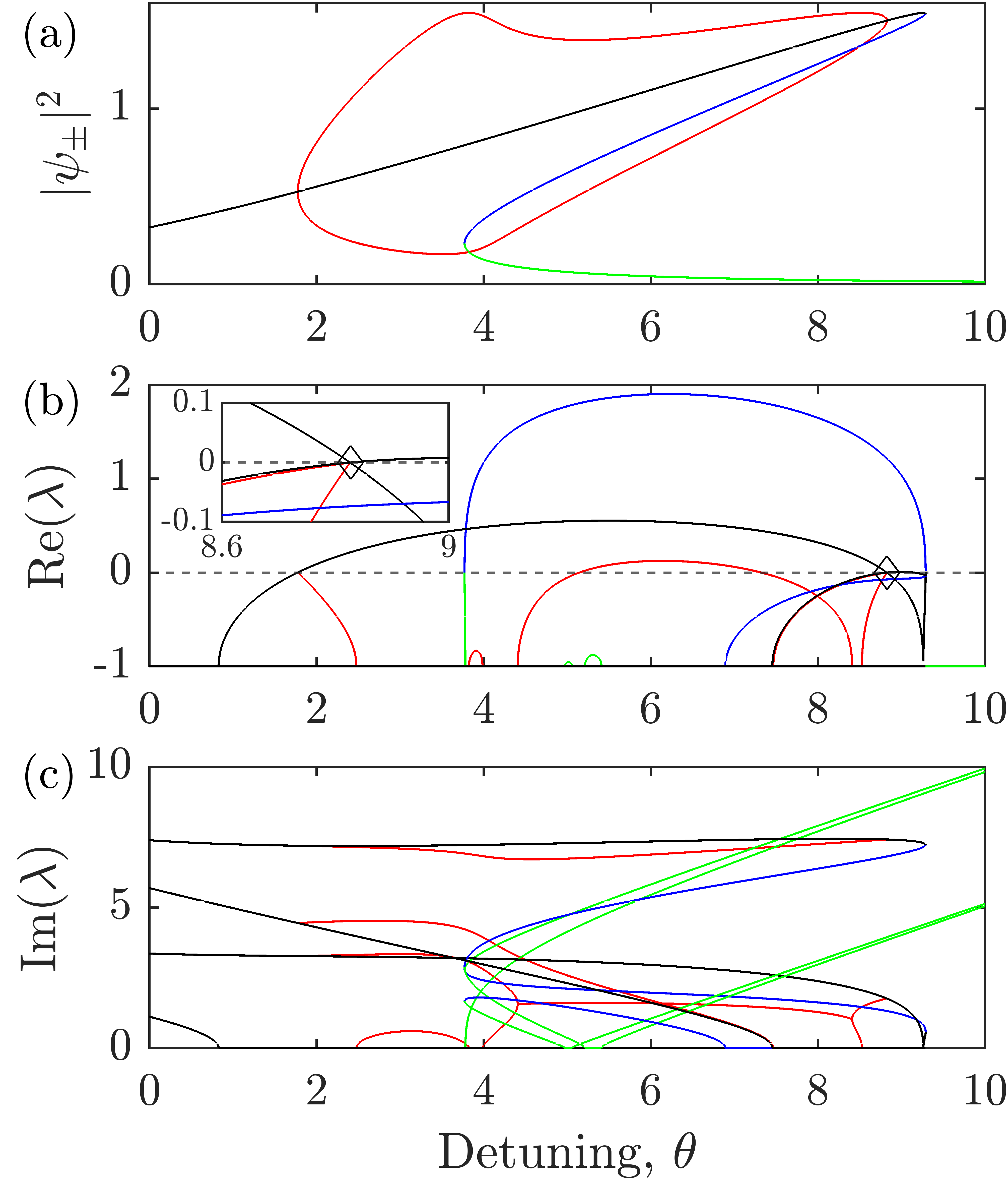}
    \caption[Linear stability eigenvalues of the homogeneous stationary solutions of the Fabry-P\'erot]{HSS (a) with the corresponding real (b) and imaginary (c) parts of the linear stability eigenvalues (\ref{eq:FPtwopolEigenvalues}) where $S = 1.2425$, for both $\alpha^2_0 = 0$ and $\alpha^2_\mu = 4.83$. Black, blue, and green curves correspond to the high, middle, and low power symmetric HSS respectively. The red curve corresponds to the symmetry broken HSS. The zoomed in window in (b) shows the codimension-2 point $(S, \theta)_{\text{codim} 2}\approx (1.2425, 8.827)$ on the high power HSS (black curves) where the real part of a $\mu = 0$ and a $\mu \neq 0$ eigenvalue are simultaneously zero, which is indicated with a black diamond.}
    \label{fig:FPtwopoleigenvalue}
\end{figure}

In Fig. \ref{fig:FPtwopoleigenvalue}(a), we plot the HSSs of Eqs. (\ref{eq:PFtwopolHSS}) for $S = 1.2425$. Their corresponding eigenvalues (Eq. (\ref{eq:FPtwopolEigenvalues})) are plotted in Fig. \ref{fig:FPtwopoleigenvalue}(b)-(c) and are evaluated for $\mu = 0$ ($\alpha^2_0 = 0$) and $\alpha^2_\mu = 4.83$. Here, we selected the value of $\alpha^2_\mu$ to be the first wavenumber with an unstable eigenvalue, corresponding to at least one eigenvalue of (\ref{eq:FPtwopolEigenvalues}) with positive real part, on the high power symmetric HSS when decreasing the detuning. This yields a total of eight eigenvalues, whose real and imaginary parts are plotted in Fig. \ref{fig:FPtwopoleigenvalue}(b)-(c), respectively. As mentioned earlier, these eigenvalues predict the results expected in this system for $\mu = 0$, where the middle branch of the tilted Lorentzian is unstable, while the lower branch is stable, and the high power HSS is unstable between the SSB pitchfork bifurcations located at
\begin{equation}
    H_\text{SSB} = \frac{1}{6} \left( 2\theta \pm \sqrt{\theta^2 - 3} \right),\label{eq:HSSbifrupoint}
\end{equation}
marking the beginning and the end of the symmetry broken HSS `bubble'. We also see a region of Hopf instability ($5.15< \theta <7.32$) for which the symmetry broken HSS are unstable to slow time oscillation. In Fig. \ref{fig:FPtwopoleigenvalue}(b)-(c) we see several so called 'exceptional points' where real eigenvalues become complex in nature \cite{Hill25}. For us these points are anything but exceptional since the stability of interest is ruled by the real part of the eigenvalues becoming positive as known since the early work of Maxwell \cite{Maxwell_1868}.

For $\mu \neq 0$, we see the appearance of an instability on the high power HSS near the peak of the Lorentzian curve. As we will see in the next section, this instability predicts the SSB of the HSS leading to the formation of Turing patterns of alternating polarization components over the fast time variable. For $S \approx 1.2425$, the appearance of the Turing instability perfectly coincides with the reverse pitchfork bifurcation of the SSB HSS bubble. At this point, shown as a diamond in Fig. \ref{fig:FPtwopoleigenvalue}(b), the real part of the relevant $\lambda_\mu(\alpha_0=0)$ and $\lambda(\alpha_\mu\neq0)$ eigenvalues are simultaneously zero and represents the transition of Re$[\lambda(\alpha_0=0)]$ form positive to negative while Re$[\lambda(\alpha_\mu\neq0)]$ goes form negative to positive. This is an example of a codimension-2 bifurcation point where two bifurcations occur simultaneously in the parameter space. In our case, both bifurcations correspond to a SSB.

\subsection{Spontaneous symmetry breaking codimension-2 bifurcation}
\begin{figure*}
    \centering\includegraphics[width=1\linewidth]{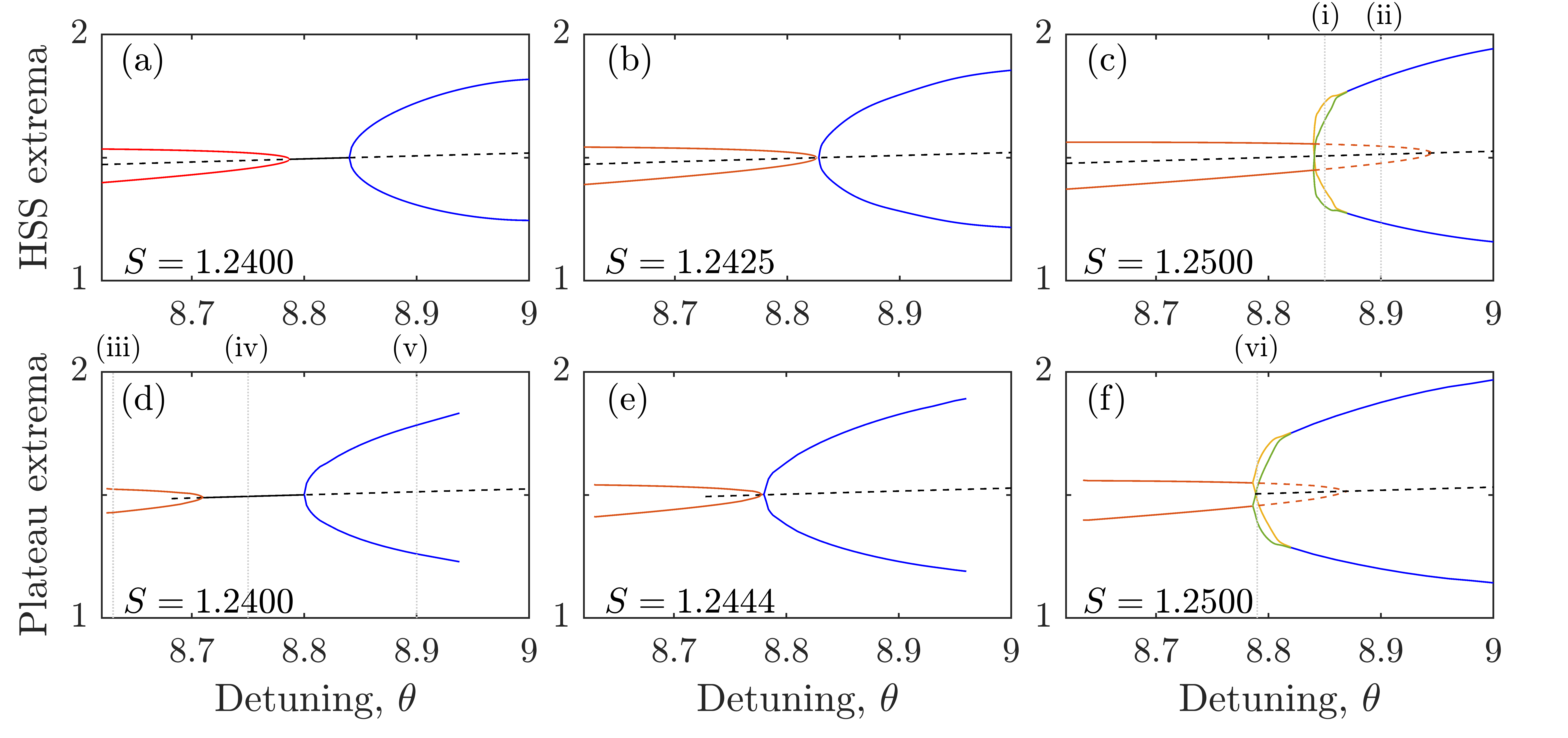}

    \includegraphics[width=1\linewidth]{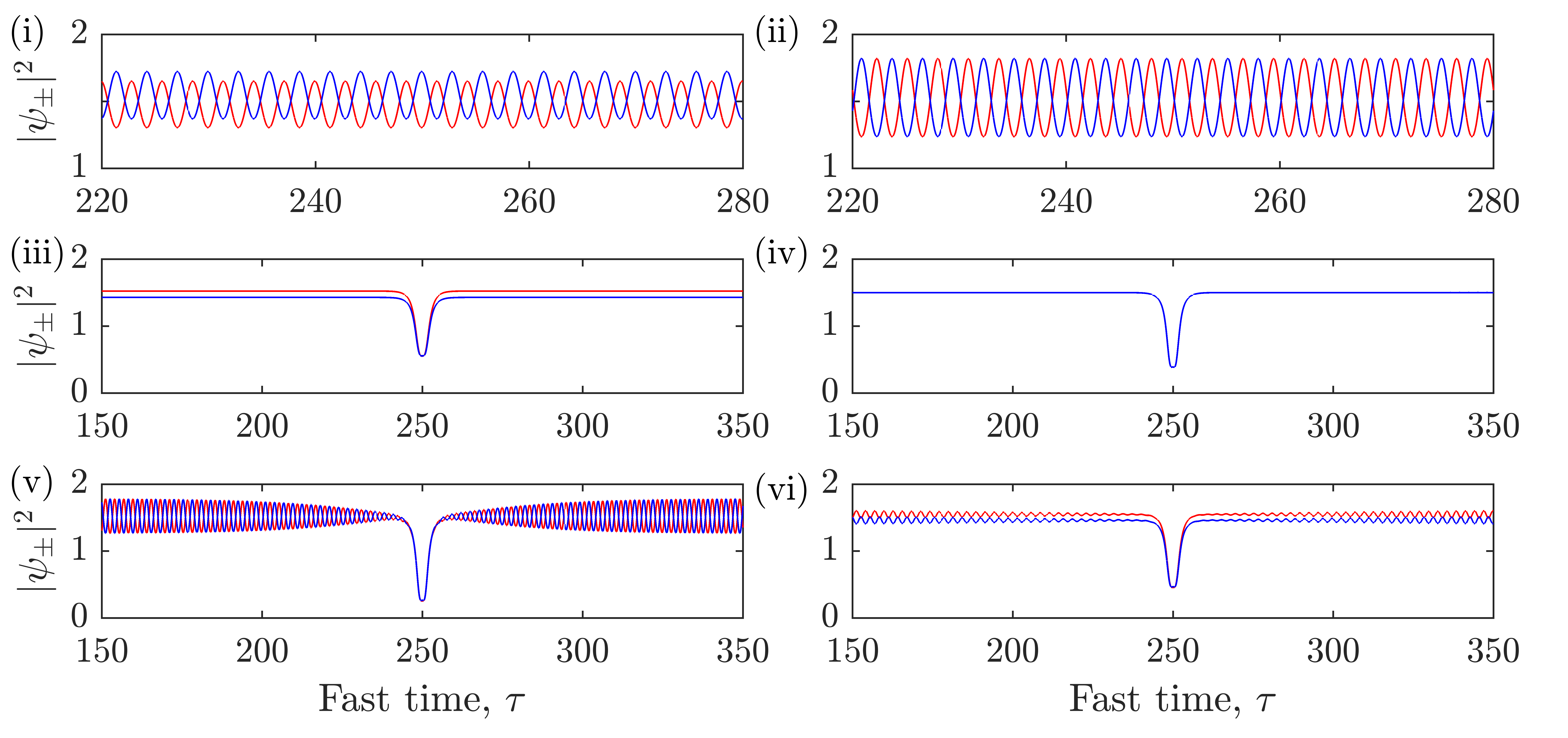}
    \caption[Codimension-2 bifurcation of the homogeneous and vectorial dark soliton stationary states]{Spontaneous symmetry bifurcations of the HSSs (a)-(c) and vectorial dark solitons (d)-(f) for $\tau_\text{R}=500$. Black curves correspond to the power of the upper symmetric HSS (a)-(c) or high power plateau (d)-(f). Red curves correspond to the power of symmetry broken HSSs (a)-(c) or symmetry broken plateaus (d)-(f). Blue, green and orange curves correspond to the maximum/minimum power of Turing patterns of alternating orthogonal polarization which form on the HSSs (a)-(c) or on the plateaus (d)-(f). (b) and (e) show a codimension-2 bifurcation at $\theta \approx 8.827$ and $\theta \approx 8.779$, respectively. (a) and (d) show the bifurcation structure for input field lesser than the codimension-2 value, $S = 1.24$,  whereas (c) and (f) show the bifurcation structure for input field amplitude $S$ greater than the codimension-2 value, $S = 1.25$. Example symmetry broken Turing patterns are shown in (i)-(ii) and vectorial dark solitons in (iii)-(vi) for $\tau_\text{R}=500$. (i) Turing pattern with unequal amplitudes for $(S,\theta) = (1.25,8.85)$. (ii) Turing pattern with equal amplitudes for $(S,\theta) = (1.25,8.9)$. (iii) Symmetry broken vectorial dark soliton with split plateaus and flat profile for $(S,\theta) = (1.24,8.63)$. (iv) Symmetric vectorial dark soliton with flat plateaus for $(S,\theta) = (1.24,8.75)$. (v) Symmetry broken vectorial dark soliton with Turing patterns on the plateau for $(S,\theta) = (1.24,8.9)$. (vi) Symmetry broken vectorial dark soliton with split plateaus and Turing patterns for $(S,\theta) = (1.25,8.79)$.}
    \label{fig:SSBofHSSandVDS}
\end{figure*}

SSB bifurcations of the high power HSS are shown in Fig. \ref{fig:SSBofHSSandVDS}(a)-(c) for different values of the input field $S$. In each of these cases the HSSs (see Eqs. (\ref{eq:PFtwopolHSS})) are plotted near the peak of the symmetric HSS Lorentzian curve, where the high power branch is shown as the black curve, and the symmetry broken HSS as the red curves. We indicate the stability of the HSS using the linear stability eigenvalues of the previous section, Eq. (\ref{eq:FPtwopolEigenvalues}), and show regions of stable and unstable HSS by solid and dashed lines, respectively. 

Fig. \ref{fig:SSBofHSSandVDS}(a) shows a region of stable symmetric HSS (solid black line) flanked by two oppositely oriented supercritical SSB bifurcations. For lower values of the detuning we see a reverse pitchfork (red curve) in which the high power symmetric HSS become unstable to symmetry broken HSSs. This is predicted by the linear stability analysis (with $\mu=0$), where the higher power symmetric HSS is unstable to two SSB pitchfork branches at the value given by Eq. (\ref{eq:HSSbifrupoint}). Of particular interest is the appearance of a bifurcation of the symmetric HSS specifically associated with the normal dispersion regime (see the blue lines in Fig. \ref{fig:SSBofHSSandVDS}(a)). This bifurcation occurs when increasing the detuning such that the symmetric HSS becomes unstable to perturbations of nonzero wavenumber ($\mu\neq0$), and results in the formation of a stationary Turing pattern of alternating polarization field components. The blue curve of Fig. \ref{fig:SSBofHSSandVDS}(a) depicts the maximum and minimum powers of the amplitude of the resulting symmetry broken Turing pattern, an example of which can be seen in Fig. \ref{fig:SSBofHSSandVDS}(ii). This instability has its origin in the interaction of local self- and cross-Kerr interaction terms of Eqs. (\ref{eq:FP}), and has previously been reported in ring resonator systems containing orthogonal polarization components \cite{campbell2024frequency}. This instability is not observed in the bidirectionally driven ring resonator \cite{campbell2022counterpropagating} due to nonlocal cross-Kerr interaction originating from counterpropagation of the fields (hence an absence of local cross-Kerr interaction).

By increasing the input field, $S$, we see that the separation between the two oppositely oriented supercritical bifurcations decreases until they meet. The critical parameter value $S \approx 1.2425, \theta\approx 8.827$ represents a codimension-2 bifurcation, where the real part of one of the $\lambda(\alpha_0=0)$ eigenvalues transitions from positive (unstable) to negative (stable), and one of the $\lambda(\alpha_\mu\neq0)$ eigenvalues transitions from negative (stable) to positive (unstable), when increasing the detuning, as shown in Fig. \ref{fig:SSBofHSSandVDS}(b). When further increasing the input field to $S = 1.25$, we see that the two bifurcations display a cross over region where symmetry broken HSSs are unstable to the formation of Turing patterns (see Fig. \ref{fig:SSBofHSSandVDS}(c)). This results in a small range of detuning values where the Turing pattern of one fields is dominant (orange curve) and the other one is submissive (green curve), an example of which can be seen in Fig. \ref{fig:SSBofHSSandVDS}(i). Although there exists a mathematical relation between HSSs in ring \cite{campbell2024frequency} and FP resonators, the ring resonator system does not display a codimension-2 bifurcation of the kind described here. This is due to different eigenspectra between the two models with respect to the formation of Turing patterns, originating from the nonlocal coupling of the FP configuration.

\section{Spontaneous symmetry breaking of vectorial dark solitons}\label{sec:FPSSB_VDS}
We first consider symmetric VDS solutions $\psi_+ = \psi_- = \psi$. At symmetry Eqs. (\ref{eq:FP}) reduce to the single field Fabry-P\'erot model under suitable renormalization  (see Eq. (\ref{eq:symFPtwopol})), and as a result, the stationary solutions of Eqs. (\ref{eq:FP}) are equivalent to the stationary states of the FP with a single field. In the numerical simulation of Eqs. (\ref{eq:FP}), we observe the formation of VDS stationary solutions in which a dark soliton is present in both field components of orthogonal polarization occupying the same fast time domain. Much like the dark solitons of the single field FP, the VDSs are themselves composed of oppositely oriented switching fronts (SFs) which connect two flat solutions (plateaus) and then lock due to their interaction through local fast time oscillations close to the lower power plateau. This mechanism was first proposed for spatial solitons composed of diffractive switching fronts \cite{rosanov1990diffractive,oppo1999domain,oppo2001characterization}, and then demonstrated longitudinally in the single field FP \cite{campbell2023dark} and ring resonator \cite{parra2016dark}. 
It should be noted that the homogeneous background plateau from which the DCSs hang does not correspond to the HSSs given by Eq. (\ref{eq:PFtwopolHSS}). The existence and stability of the plateaus depend on the average power of the intracavity field over a round trip. This is a result of the nonlocal coupling terms originating form the counterpropagation of intracavity fields, and has also been demonstrated for bidirectionally pumped ring resonators \cite{campbell2022counterpropagating}. 

At symmetry, the plateau solutions can be approximately calculated by solving the coupled equations
\begin{eqnarray}
    4Y_{u,l}^3 &-& 4(\theta - 4\Delta Y_l - 4(1-\Delta)Y_u)Y_{u,l}^2\label{eq:upperandlowerPlateau}\\
    &+& \{(\theta - 4\Delta Y_l - 4(1-\Delta)Y_u)^2 + 1\}Y_{u,l} = S^2,\nonumber
\end{eqnarray}
where $\psi$ is assumed to be composed of upper and lower plateaus connected by step functions. Here, we define $Y_{l}$ as the power of the lower plateau and $Y_{u}$ as the power of the upper plateau, such that the average power over the round trip can be expressed as $\Delta Y_l + (1-\Delta)Y_u$, where $0<\Delta<1$ is the normalized duration of the lower plateau over the cavity round trip time. This assumption has been used to great effectiveness in predicting the stationary solutions is related systems \cite{campbell2022counterpropagating,campbell2023dark}.  

In Figs. \ref{fig:SSBofHSSandVDS}(d)-(f), we show the SSB bifurcations of the background plateau for a single VDS present in the cavity when scanning the detuning $\theta$ for different values of the amplitude of the input field $S$. The plateaus of symmetric VDSs display a SSB phenomenon analogous to that observed for HSSs, but are now affected by the presence of nonlocal coupling. In Fig. \ref{fig:SSBofHSSandVDS}(d) we plot the plateau power of a symmetric VDS as a black curve while an example of a stable symmetric VDS is shown in Fig. \ref{fig:SSBofHSSandVDS}(iv). Here, regions of stable and unstable VDSs are indicated by solid and dashed lines, respectively. Similar to the corresponding HSS, stable symmetric VDSs are flanked by two oppositely oriented SSB bifurcations. If we decrease the detuning we encounter a reverse pitchfork bifurcation at $\theta \approx 8.71$ in which one dark soliton hangs from a higher power plateau and the other hangs from a lower power plateau, as shown in Fig. \ref{fig:SSBofHSSandVDS}(iii). The power of the dominant and submissive plateaus are plotted as red and blue curves, respectively, and were determined numerically. This SSB bifurcation is associated with the reverse pitchfork bifurcation of the HSSs, although the bifurcation point is now displaced with respect to the case of HSSs. This is a consequence of the shift in detuning originating from counterpropagation $\theta_{\text{eff}} = \theta - 4\langle|\psi|^2\rangle$ induced by the presence of the VDS \cite{cole2018theory,hill2024symmetry,campbell2022counterpropagating,campbell2023dark,Fan20}. In the limit of large cavity round trips, we have that $\langle|\psi_\pm|^2\rangle\rightarrow H_\pm$. As a result, in this limit the bifurcation point of the plateau approaches the bifurcation point of the HSS. 

If we now increase the detuning to $\theta \approx 8.8$, we encounter a supercritical SSB bifurcation of the VDSs resulting in the formation of a fast time Turing pattern of alternating polarizations on the background plateau of the VDS. The blue curve of Fig. \ref{fig:SSBofHSSandVDS}(d) depicts the maximum and minimum power of the amplitude of the resulting symmetry broken Turing pattern. After this bifurcation point, the symmetric VDS branch becomes unstable. An example of a symmetry broken stable VDS solution is shown in Fig. \ref{fig:SSBofHSSandVDS}(v) showing rapid fast time oscillations of the background power between the fields of orthogonal polarization whose amplitude decreases on approach to the VDS. The formation of symmetry broken Turing patterns has been demonstrated earlier for the HSSs, where the Turing instability can be similarly understood as originating form the contributions of local self- and cross-Kerr interactions of Eqs. (\ref{eq:FP}). Again, we see that the bifurcation point has been displaced with respect to the equivalent instability of the HSSs.

In Fig. \ref{fig:SSBofHSSandVDS}(e), we show the SSB bifurcations of the VDSs for $S = 1.2444$. At the values $(S,\theta)_\text{codim 2} \approx (1.2444,8.78)$, we find a codimension-2 bifurcation, where the reverse pitchfork bifurcation and SSB Turing pattern bifurcation occur at exactly at the same value of the detuning. The codimension-2 point of the VDS occurs at a different place of parameter space when compared to the HSSs due to the shift detuning introduced by the integral terms in the model equations. In general, the location and existence of these two SSB bifurcations depend on the average power (size and number of solitons within the cavity) of the symmetric VDS solution. Further increasing the input field to $S=1.25$, we see in Fig. \ref{fig:SSBofHSSandVDS}(c) that the SSB bifurcations have now crossed over leading to regions of parameter space in which one pattern component is dominant and the other submissive, such as in the case shown in Fig. \ref{fig:SSBofHSSandVDS}(vi).

The formation of Turing patterns via SSB of the symmetric VDSs can be investigated by approximating the VDSs as two plateaus occupying distinct domains of the fast time variable and connected by step functions $\psi^\text{s} = \psi^\text{s}_{u} + \psi^\text{s}_{l}$, where $\psi_{u,l}^\text{s}$ are the plateau solutions of higher $u$ and lower $l$ powers. We introduce a linear perturbation to each plateau of the form 
\begin{align}
    \psi_{\pm,u}(\tau,t) &= \psi_u^\text{s} + \epsilon_{\pm,u} e^{i\alpha_u \tau+\Omega_u t}\label{eq:FP2polPlatPertu}\\
    \psi_{\pm,l}(\tau,t) &= \psi_l^\text{s} + \epsilon_{\pm,l} e^{i\alpha_l \tau+\Omega_l t}\label{eq:FP2polPlatPertl}
\end{align}
where $\alpha_{u,l}$ and $\Omega_{u,l}$ are the wavenumbers and growth rates of the perturbation on the respective plateaus, and $\epsilon_{\pm,u}, \epsilon_{\pm,l}<<1$. Inspired by the linear stability analysis previously performed for the HSSs, we assume that the perturbation does not survive the integral terms, such that perturbations (\ref{eq:FP2polPlatPertu}) and (\ref{eq:FP2polPlatPertl}) obey $\langle |\psi_+|^2\rangle = \langle |\psi_-|^2\rangle = \langle \psi_+\psi_-^*\rangle = \langle \psi_+^*\psi_-\rangle =  \langle |\psi^\text{s}|^2\rangle = \Delta (\psi_l^s)^2 + (1-\Delta)(\psi_u^s)^2$. This is akin to assuming that the wavenumbers $\alpha_{u,l}$ are periodic on their respective plateau. As the perturbations do not survive the nonlocal coupling terms, perturbations on the upper and lower plateau are no longer coupled, so that, without loss of generality, we can consider the high and lower power plateaus to be real. Inserting the step function approximation into Eqs. (\ref{eq:FP}) and following the procedure outlined in Appendix \ref{app:stabplat}, we arrive at the linear stability eigenvalues 
\begin{align}    
    \Omega_{u,l}(\alpha_{u,l}) = -1 \pm \bigg[&-A_{u,l}B_{u,l} -C_1C_2\nonumber\\
    &\pm (A_{u,l} C_2 + B_{u,l} C_1)\bigg]^{1/2},\label{eq:pat_eigenvalues1}
\end{align}
where $A_{u,l} = \theta-\alpha_{u,l}^2 - 2(\psi_{u,l}^s)^2 - 8/3\langle|\psi^s|^2\rangle$, $B_{u,l} = \theta-\alpha_{u,l}^2 - 10/3(\psi_{u,l}^s)^2 - 8/3\langle|\psi^s|^2\rangle$, $C_1 = - 4/3\langle|\psi^s|^2\rangle$, $C_2 = C_1 - 8/3(\psi_{u,l}^s)^2$, ($\psi_{u,l}^s)^2$ are the powers of the high and lower power plateaus (here considered to be real) and $\langle|\psi^s|^2\rangle$ is the average of the intracavity field. We have performed similar linear stability analysis in the presence of integral terms to good effectiveness in \cite{campbell2023dark,campbell2022counterpropagating} and we find that this linear stability analysis accurately predicts the onset of pattern instability for symmetric VDSs too. We note that the eigenvalues of Eqs. (\ref{eq:pat_eigenvalues1}) reduce to the eigenvalues of symmetric HSS in the limit $\langle|\psi^s|^2\rangle\rightarrow|\psi^s|^2$, further cementing the connection with the HSSs.

\section{Spontaneous symmetry breaking in the presence of nonlocal coupling}
\begin{figure}
    \centering\includegraphics[width=1\linewidth]{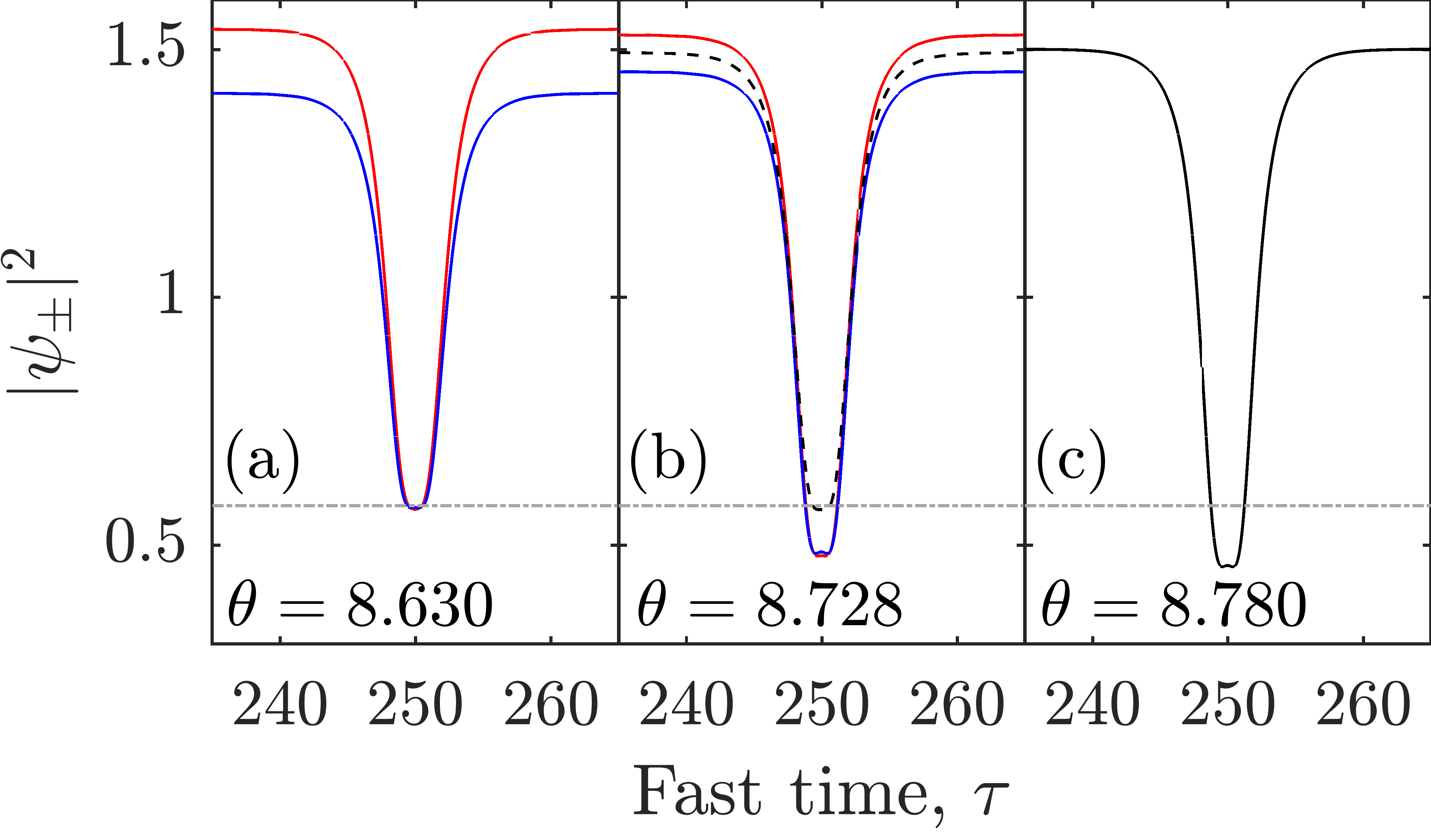}

    \includegraphics[width=1\linewidth]{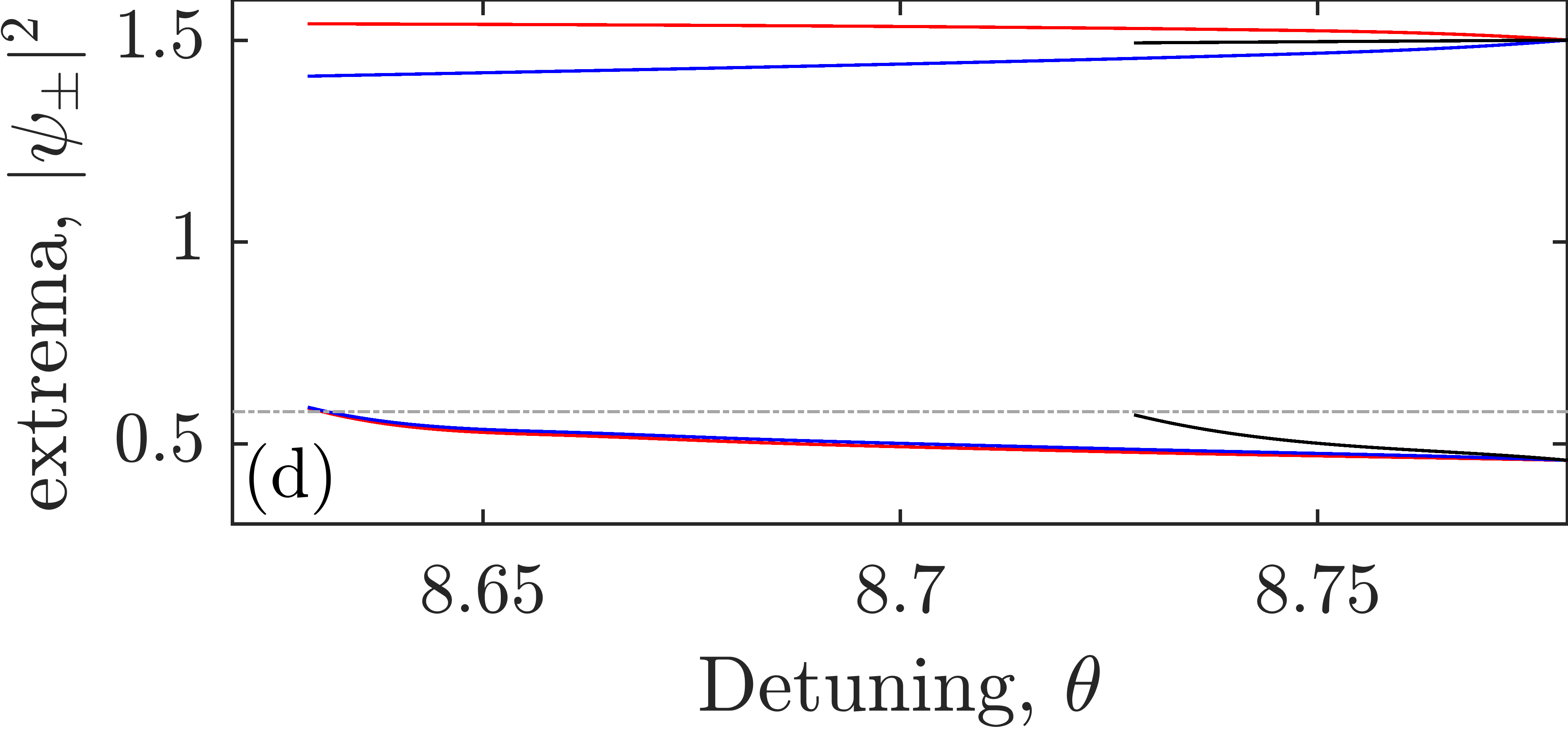}
    \caption[Extended range of existence of symmetry broken dark solitons]{(a) The VDS solution (black curve) at the codimension-2 point $(S, \theta)_\text{codim 2} \approx (1.2444, 8.78)$. By decreasing the detuning, we observe SSB of the VDS where the symmetric solution becomes unstable (dashed black curve) to the symmetry broken VDS (red and blue curves). (b) corresponds to the saddle node bifurcation which marks the end of the of the symmetric VDS solution branch. Further decreasing the detuning, we see the unstable symmetric VDS solution has disappeared such that only the symmetry broken VDS remains (c). The plateau power and minimum power of the symmetric (black) and symmetry broken (red and blue) VDS is shown in (d) when scanning detuning.}
    \label{fig:deathofsym}
\end{figure}
In the bifurcation diagrams of Fig. \ref{fig:SSBofHSSandVDS}(d)-(f), it can be seen that the symmetry broken VDS branches (red curves) exist far beyond the end of the symmetric (now unstable) soliton branch, when decreasing the detuning $\theta$. This phenomenon is atypical for SSB where the symmetric solution usually persist over the full domain of the symmetry broken solutions, as seen for the HSSs in Fig. \ref{fig:FPtwopoleigenvalue}. In Fig. \ref{fig:deathofsym}(c) we plot a single VDS present at the codimension-2 point $(S,\theta)_\text{codim 2} \approx (1.2444,8.78)$. By decreasing the detuning, we observe the SSB of the VDS. The power at the trough of the unstable symmetric VDS increases as the the detuning is decreased until we reach a saddle node bifurcation, marking the end of the symmetric VDS solution branch. The final symmetric VDS is shown with a dashed line in Fig. \ref{fig:deathofsym}(b). The interaction of the oscillatory tails of the SFs, to which the VDSs owe their existence is highly dependent on the power of the lower plateau close to the soliton trough. This in turn is highly relevant in determining the decay rate and wavenumber of these oscillations. Beyond this value of detuning, the interaction of the oscillatory tails is insufficient to induce the locking mechanism due to the increasing power of the trough. As can be seen in Fig. \ref{fig:deathofsym}(b), the stable symmetry broken VDS has bifurcated from the symmetric solutions and maintains a lower value of trough power. By further decreasing the detuning, the power of the trough of the symmetry broken VDS rises until the locking mechanism fails for it as well as for the symmetric VDS as shown in Fig. \ref{fig:deathofsym}(d). This failure of the locking mechanism is delayed for the symmetry broken solutions due to the slower rate of increase in the trough power of the VDSs. This phenomenon is a result of the nonlocal coupling, whose contributions differ between symmetric and symmetry broken solutions, and acts to displace the existence of the symmetry broken solutions beyond the range of the symmetric solution branch from which they bifurcate. The final result is an extended range of parameter values in which only symmetry broken VDSs exist.


\section{Self-crystallization of vectorial dark solitons}\label{sec:FPSSB_selfcry}
\begin{figure*}
    \centering
    \includegraphics[width=1\linewidth]{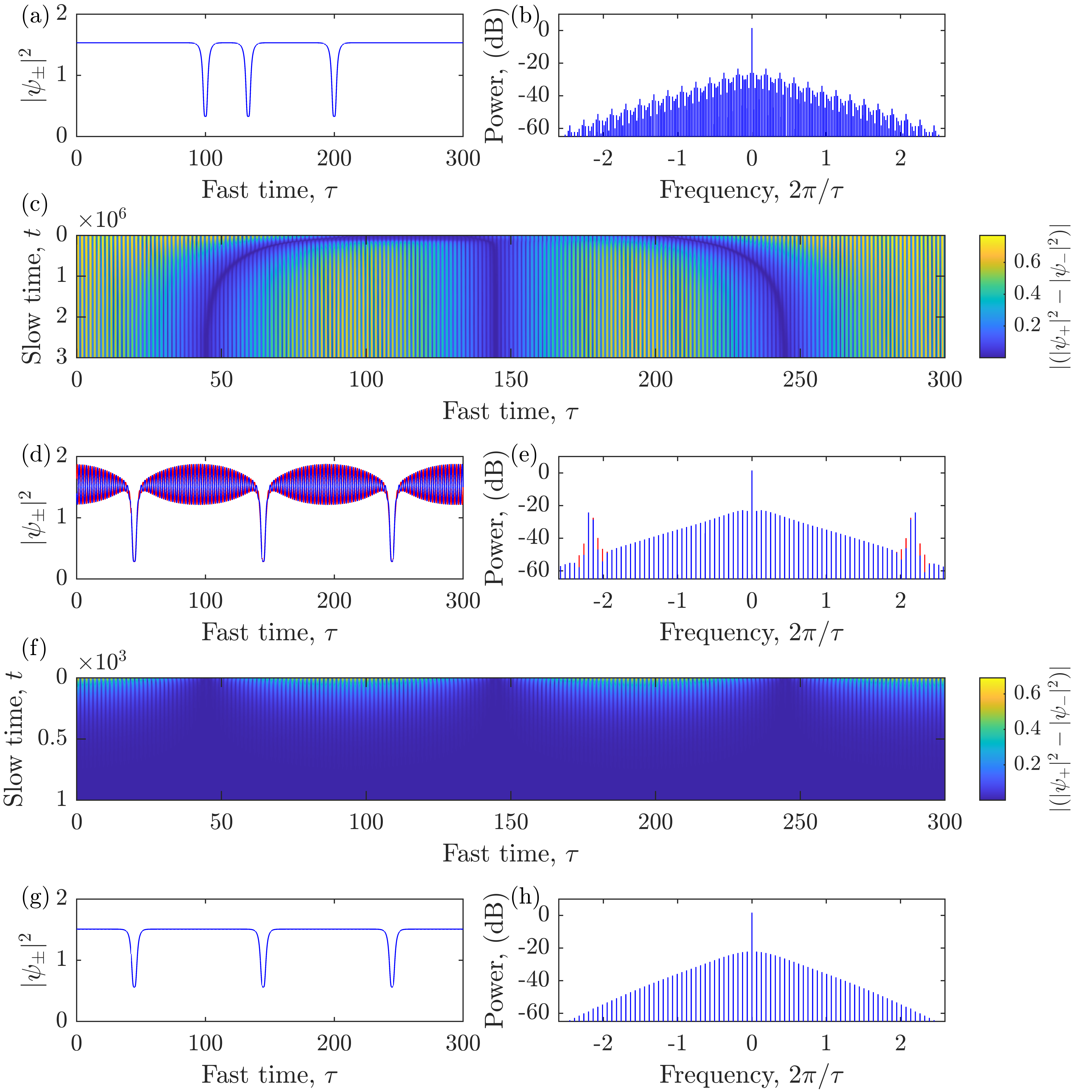}
    \caption[Self-crystallisation of three vectorial dark solitons in Fabry-P\'erot resonator.]{(a) Initial condition of three symmetric vectorial dark solitons and corresponding comb spectrum (b). (c) Slow time evolution of the initial condition in (a) for $S=1.25$, $\theta = 8.8$ demonstrating the growth of SSB Turing patterns, which propel the VDSs through the cavity. (d) Final RSC stationary state and corresponding comb spectrum (e). (f) Slow time evolution after reducing the cavity detuning to $\theta = 8.63$, below the threshold for Turing patterns. (g) Final regular VDS crystal with a uniform background and corresponding comb spectrum (h).}\label{fig:FPselfcry}
\end{figure*}
We now consider a cavity containing multiple VDSs along the round trip time. These solutions undergo the SSB bifurcations discussed earlier, such that Turing patterns form in the intervals between adjacent VDSs. The growth of Turing patterns induces motion in the VDSs resulting in adjacent VDSs to be ‘pushed’ apart until an equilibrium of the pattern’s amplitude is reached on both sides of each VDS. Hence, the SSB phenomenon in the plateaus around VDSs introduces long range repulsive interaction between adjacent VDSs. In the absence of symmetry broken Turing patterns the symmetric VDSs do not exhibit long range repulsive interactions and they remain stationary at arbitrary separation distances. 

This process is analogous to the `self-crystallization' phenomenon described for the ring resonator with vectorial intracavity fields in \cite{campbell2024frequency}. In the ring resonator, the long range interaction between solitons can only be mediated by the Turing patterns, as nonlocal coupling is absent in that case. Here, we demonstrate that the `self-crystallization' mechanism can be generalized to FP resonators with orthogonal polarizations components too \cite{Campbell2023fabryperot2pol}, and as such, it is universal in systems displaying temporal cavity solitons and Turing instabilities under local self- and cross-phase modulation. We also discuss the differences in self-crystallization between ring and FP resonators and the effects originating form nonlocal coupling.

\subsection{Spontaneous formation of regularly spaced soliton crystals}

In Fig. \ref{fig:FPselfcry}(a), we introduce three symmetric VDSs randomly distributed along the cavity round trip. For the parameter values $S=1.25$, $\theta = 8.8$, $\tau_\text{R} = 300$, the background plateau that supports the VDSs is unstable to the formation of symmetry broken Turing patterns. The maximum amplitude reached by the Turing patterns in the intervals between the VDSs depends on the separation of adjacent VDSs. As the pattern amplitude grows, the VDS are ‘pushed’ along the resonator until an equilibrium configuration of the pattern is reached on either side of each VDS. In Fig. \ref{fig:FPselfcry}(c), we show the slow time evolution of three VDSs. Here it can be seen that the VDSs move as to spread out along the cavity coordinate which results in the stable stationary state shown in Fig. \ref{fig:FPselfcry}(d). This stationary solution is composed of VDSs located equidistantly along the round trip time of the cavity, separated by Turing patterns of equal amplitude forming a perfectly regular soliton crystal (RSC).

The formation of a RSC from the initial condition of three randomly positioned VDSs is spontaneous, induced by SSB of the VDSs, and is an example of a `self-crystallization' of a random distribution of VDSs \cite{campbell2024frequency}. This spontaneous self-organization of the VDSs is induced by the long range interactions originating from the SSB of the intracavity field and does not rely on any perturbations introduced to the mode equations, such as those introduced near avoided mode crossings \cite{cole2017soliton}, or to describe external modulations of the driving \cite{lu2021synthesized}. Furthermore, by moving the detuning back across the Turing instability bifurcation, background pattern modulations in the intervals between VDSs vanish, as shown in Fig. \ref{fig:FPselfcry}(f). This evolution results in the formation of a symmetric RSC of Fig. \ref{fig:FPselfcry}(g). While the long range interaction resulting from the nonlocal (global) coupling is always present in the FP resonator, self-crystallization relies on the interactions of VDSs mediated by the Turing patterns, and as such, VDSs will not display self-crystallization phenomena below the SSB Turing pattern bifurcation.

The RSC created by this self-organization phenomenon produces a frequency comb (see Fig. \ref{fig:FPselfcry}(e)) with a smooth spectral envelope and a free spectral range three times larger than the frequency comb of the initial condition, Fig. \ref{fig:FPselfcry}(b). Regular peaks appear in the spectral envelope due to the contribution of the Turing pattern wavenumber to the spectral envelope, that is required for self-crystallization. Such peaks can be removed at will after self-crystallization by changing the control parameters back below the SSB bifurcation, as shown in \ref{fig:FPselfcry}(f), thus leaving a symmetric RSC with no pattern states between the VDS as shown in Fig. \ref{fig:FPselfcry}(g) and (h). In general, a RSC composed of $N$ VDSs produces a frequency comb equivalent to a single VDS in a cavity with round trip $\tau_\text{R}/N$. An important property of RCSs is their capacity to emulate smaller cavity sizes via an increasing soliton numbers, leading to frequency combs with enhanced power and greater spacing of the spectral lines. As such the formation of RSC has many potential applications, such as satellite communications \cite{federici2010review}, photonic radar \cite{riemensberger2020massively} and radio-frequency filters \cite{xu2018advanced,hu2020reconfigurable}, where the universality of this mechanism in ring and FP resonators creates different practical methods for the generation and control of RSCs, distinct from previously demonstrated methods \cite{cole2017soliton,lu2021synthesized}.

\subsection{Partial crystal formation and interaction length}
\begin{figure*}
    \centering    
    \includegraphics[width=1\linewidth]{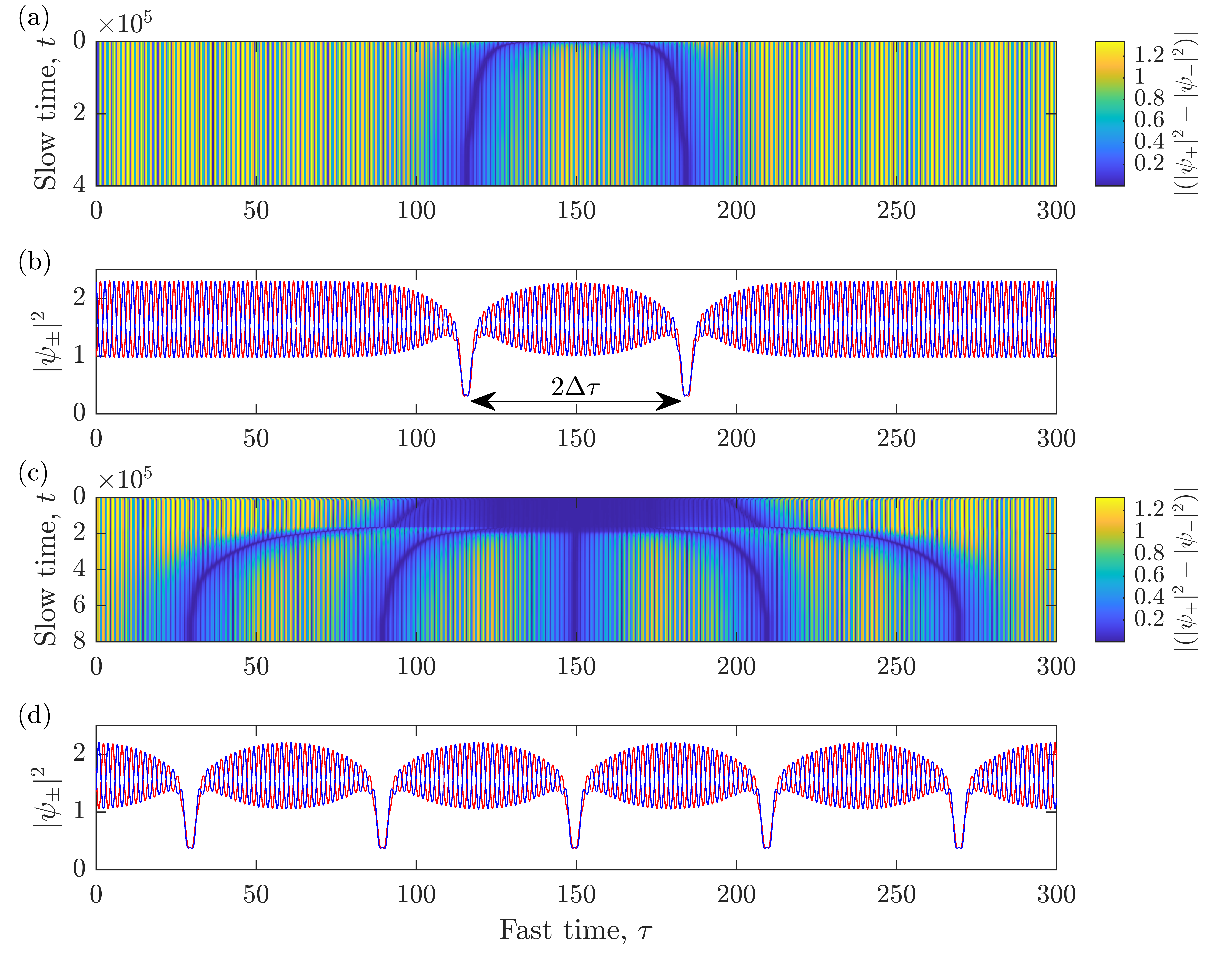}
    \caption[Partial crystal formation in Fabry-P\'erot resonator.]{(a) Slow time evolution of two VDSs resulting in the formation of the partial soliton crystal (b) with VDS spacing $2\Delta\tau$ for $S=1.05\sqrt{3/2},\theta=9.1$. (c) Slow time evolution of five VDSs form a tightly packed initial condition. This evolution results in the RSC stationary state (d) for $S=1.05\sqrt{3/2},\theta=8.9$.}
    \label{fig:FPpartialcry}
\end{figure*}
As mentioned in the last Section, adjacent VDSs can move apart due to the growth of Turing patterns in the interval between the VDSs. The motion of the VDSs is observed until the Turing patterns on both sides of the solitons reach saturation. In Fig. \ref{fig:FPpartialcry}(a), we show the evolution of two VDSs in a long cavity. This results in the formation of a partial crystal of Fig. \ref{fig:FPpartialcry}(b), composed of two (or more) equidistant VDSs in a local region of the cavity \cite{campbell2024frequency}, with the remainder of the cavity occupied by a Turing pattern. From Fig. \ref{fig:FPpartialcry}(b) we define the VDS spacing of a partial crystal as $2\Delta\tau$, i.e. twice the distance in the fast time variable to the place where the amplitude of the pattern reaches its maximum value. This characteristic length can be used to determine the condition required for self-crystallization of a RSC, $2N\Delta\tau > \tau_R$ where $N$ is the number of VDSs in the cavity, as seen in Fig. \ref{fig:FPpartialcry}(c)-(d) for five initial VDS. RSCs of the FP resonator are robust to a change in the number of VDS as the repulsive interactions will redistribute VDSs to equidistant locations. This is conditional on the Turing patterns saturation length $2\Delta\tau$ which must be less that the lattice spacing the RSC. 

In Fig. \ref{fig:FPpartialcry}(c) we show the evolution of five tightly packed VDS in a FP resonator, resulting in the formation of a RSC, Fig. \ref{fig:FPpartialcry}(d). This evolution is split into two stages. For $t<2\times 10^5$ the Turing patterns amplitudes between the tightly packed VDSs is small (near symmetry) due to the long pattern saturation distance ($\Delta\tau$), resulting in slow dynamics of the VDSs. This continues until the VDS are sufficiently far apart to support Turing patterns of larger amplitudes, at $t\approx 2\times 10^5$, where the VDS dynamics become significantly faster. If we compare with the analogous phenomenon of the ring resonator \cite{campbell2024frequency}, we find that the pattern saturation distance is larger in the FP, and as such the FP is more prone to the slow dynamics displayed in Fig. \ref{fig:FPpartialcry}(c). Conversely, the greater value of $2\Delta\tau$ suggest that the FP resonator is more congruent to the formation of RSCs, with greater robustness to changes in soliton number.   

\begin{figure}
    \centering\includegraphics[width = 1\linewidth]{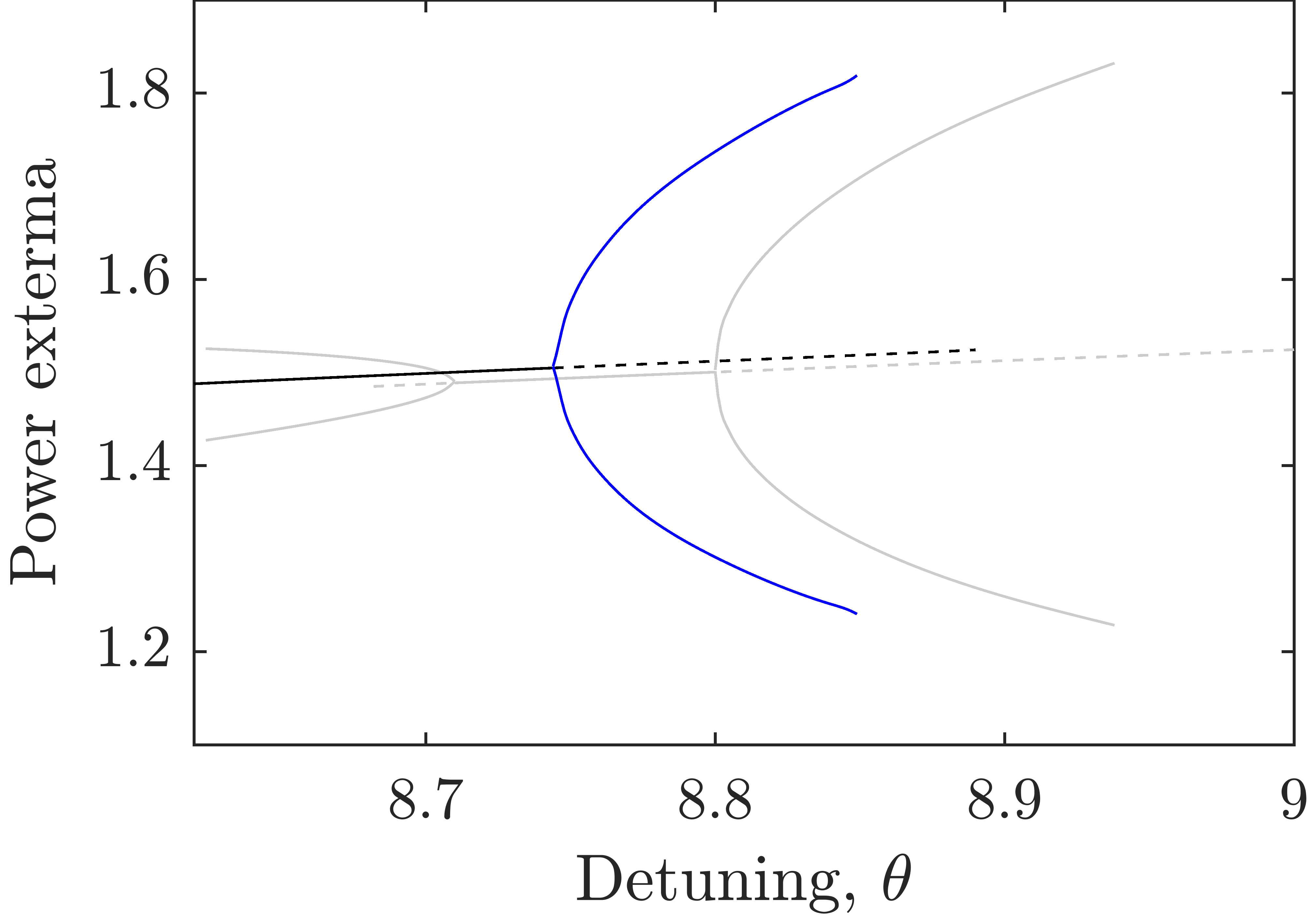}
    \caption[Displaced bifurcation of three vectorial dark solitons]{Spontaneous symmetry breaking bifurcations for $S=1.24, \tau_\text{R}=500$ and three VDSs. The plateau power of three symmetric VDSs is plotted as a black curve. The blue curve show the maximum and minimum amplitude of the symmetry broken patterns of alternating polarization which forms in between equidistant VDSs. The gray curves show the single VDS bifurcations of Fig. \ref{fig:SSBofHSSandVDS}(a). In both of these cases stable (unstable) solutions are depicted with solid (dashed) lines. This demonstrates the noticeable displacement of the locations of the bifurcation when increasing the number of solitons form one to three, where we note the absence of the reverse pitchfork bifurcation for the three soliton case.}
    \label{fig:SSBofthreeDSPs}
\end{figure}

The nonlocal coupling present in systems of counterpropagating light introduces long range interactions between well separated solitons within the cavity, such that, the existence and stability of the solitons is dependent on the soliton number \cite{campbell2023dark,campbell2022counterpropagating}. The symmetric VDSs of the two component FP, Eqs. (\ref{eq:FP}), display such long range interactions. In Fig. \ref{fig:SSBofthreeDSPs}, we plot the SSB bifurcation diagram of the plateau when there are three simultaneous VDSs in the cavity for $S=1.24$ and $\tau_\text{R}=500$. We see that the symmetric VDS solution undergoes a SSB at $\theta \approx 8.75$ resulting in the formation of a symmetry broken Turing pattern and the formation of a RSC, where maximum and minimum powers of the Turing pattern are plotted as the blue curves. For these parameter values, solutions containing three VDS may exists as either a symmetry broken regular soliton crystal for $\theta > 8.75$ or as a symmetric random distribution along the cavity for $\theta < 8.75$. In this case a soliton crystal can be formed from a random distribution of VDSs by scanning the detuning to induce SSB of the plateau leading to the self-crystallization of the VDSs. We note that solutions containing three VDSs do not possess the reverse pitchfork bifurcation typical of single VDS solutions and shown with the grey curves in Fig. \ref{fig:SSBofthreeDSPs}. 

\section{Vectorial dark-bright solitons}\label{sec:FPSSB_VDBS}

\begin{figure}
    \centering
    \includegraphics[width=1\linewidth]{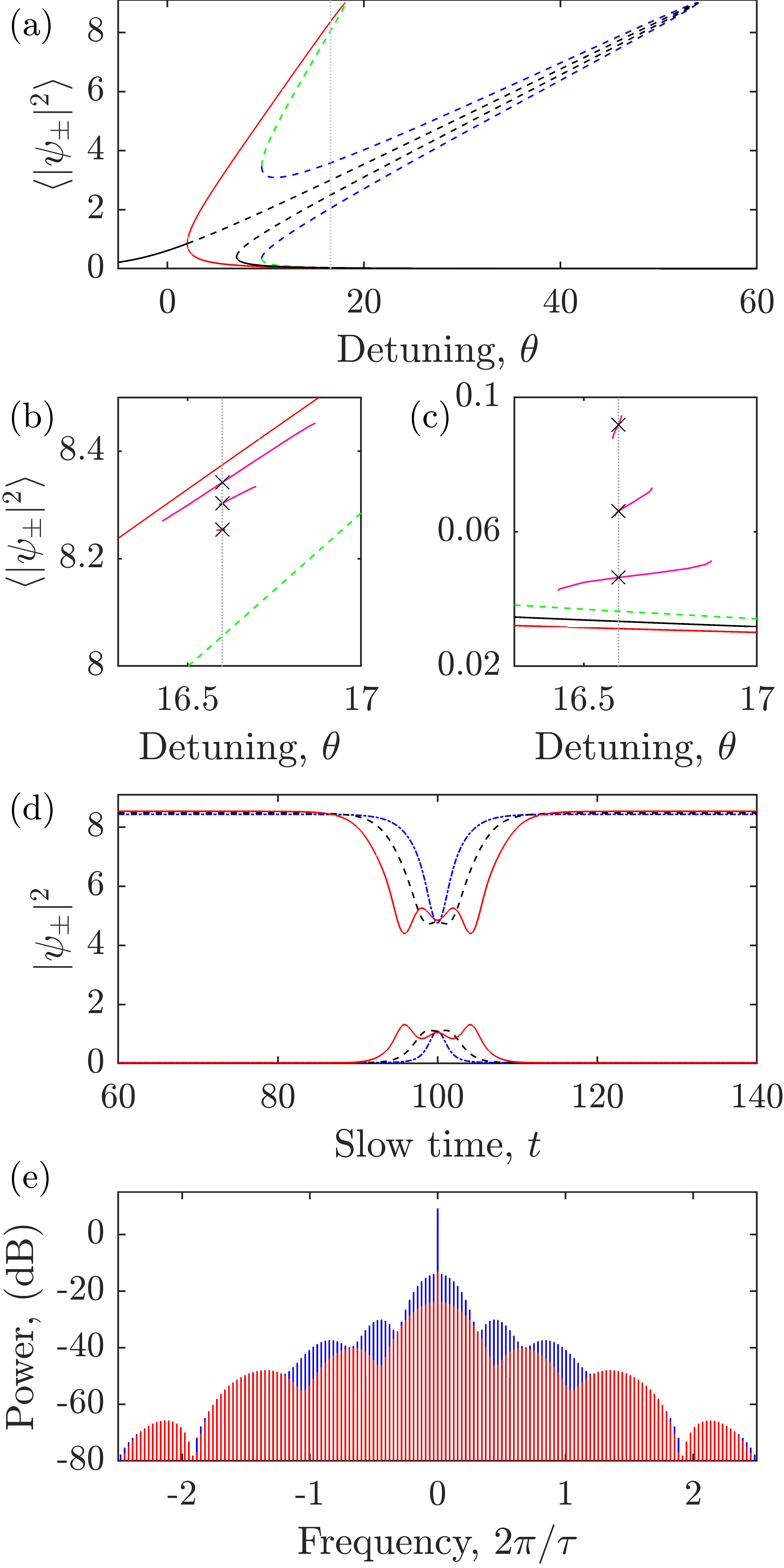}
    \caption[Dark-bright vectorial dark solitons in Fabry-P\'erot resonator]{(a) Solutions of the FP for $S=3, \tau_\text{R}=200$ are plotted as their average power over the round trip. The symmetric and symmetry broken HSS are plotted as the black and red curves respectively. In the vicinity of $\theta = 16.6$ we plot stable branches of vectorial dark-bright solitons. This collapse snaking behavior can be seen near the dominant symmetry broken HSS (b), corresponding to the dark soliton, and near the suppressed symmetry broken HSS (c), corresponding to the bright soliton. (d) Three bistable VDBSs for parameters $S=3,\theta=16.6,\tau_\text{R}=200$. Three solutions are shown, each for identical parameters, with different locking distances corresponding to one trough (dot dashed blue line), two troughs (dashed black), and three troughs (solid red line). They are indicated in (b) and (c) as x's. (d) Dual frequency comb corresponding to the dark (blue) and bright (red) solitons of the red curve in (d).}
    \label{fig:dark_bright_soliton}
\end{figure}
We now change the focus of our investigation to vectorial soliton solutions of Eqs. (\ref{eq:FP}) which exist in parameter regions displaying strongly symmetry broken HSSs. In Fig. \ref{fig:dark_bright_soliton}(a), we show the HSSs for $S=3$. A notable feature of the HSSs for this value of $S$ is the formation of a `horn' in the symmetry broken HSSs, where one polarization component is largely dominant, approaching the peak power, and the other is largely suppressed, approaching zero power (i.e. $\psi_+\gg\psi_-$). This horn gives rise to a small region of optical bistability between symmetry broken HSSs corresponding to the red and blue curves of Fig. \ref{fig:dark_bright_soliton}(a)-(c). Near the peak of the symmetry broken HSS horn, and with suitable perturbations, we observe the formation of vectorial dark-bright cavity solitons (VDBS). These solutions comprise a pair of coupled solitons occupying the same domain in fast time, where the dark soliton hangs form the dominant field and a bright soliton sits upon the suppressed low power field, as is shown in Fig. \ref{fig:dark_bright_soliton}(d).

\subsection{Stationary vectorial dark-bright solitons in Fabry-P\'erot resonators}

VDBS solutions have been demonstrated in a variety of physical systems, such as in single mode optical fibers \cite{afanasjev1989nonlinear,trillo1988optical,lisak1990symbiotic,buryak1996coupling,afanasyev1989dynamics,kivshar1991symbiotic,afanas1988nonlinear,christodoulides1988black,hu2019observation}, was well as two-species Bose-Einstein Condensates \cite{becker2008oscillations,hamner2011generation,busch2001dark}. The formation of the VDBSs in these cases is reliant on the nonlinear cross-phase modulation between the two respective field components. Despite the significantly different physics, these systems are mathematically analogous and governed by coupled nonlinear Schr\"odinger equations. In optical microresonator systems, the generation of VDBSs can be achieved though bichromatic input fields. In particular, the generation of frequency combs via bichromatically driven resonators has been widely demonstrated to produce bright-bright soliton pairs in \cite{strekalov2009generation,hansson2014bichromatically,taheri2017optical,bao2017dual,wang2016dual,weiblen2019bichromatic}, where in addition, the simultaneous generation of frequency combs between fields of orthogonal polarizations has been demonstrated \cite{suzuki2018theoretical,bao2019orthogonally}. More recently, VDBS have also been experimentally demonstrated in microresonators \cite{zhang2020Spectralextension,zhang2022dark}. This was achieved by appropriately selecting the two driving frequencies, such that, one of the intracavity fields operates within the regime of anomalous group velocity dispersion, generating a bright soliton, while the second field operates in the normal group velocity dispersion regime, which supports a dark soliton through cross-phase modulation with the bright soliton. Dark and bright solitons are bound in the fast time due to Kerr interaction and copropagate along the microresonator. 

Here, we present a different configuration for the formation of VDBSs in a FP resonator, where the generation of VDBSs of Eqs. (\ref{eq:FP}) is achieved with a monochromatic linearly polarized input field, supporting two counter-rotating polarization field components exhibiting identical normal dispersion. VDBSs of Eqs. (\ref{eq:FP}) are composed of SFs which connect to four different plateaus due to the bistability curve present over the interval of existence of the horn, as can be seen for both the dominant and suppressed fields in Figs. \ref{fig:dark_bright_soliton}(b)-\ref{fig:dark_bright_soliton}(c), respectively. In Fig. \ref{fig:dark_bright_soliton}(d), we show three multi-stable VDBSs for $S=3, \theta=18.6$. The dark and bright solitons of the VDBS are each composed of SFs that connect the two high power and two low power plateaus, respectively, within the regime of optical bistability at the horn. In the dominant field, SFs composing the dark soliton (shown as an `x' in Fig. \ref{fig:dark_bright_soliton}(b)) connect to plateaus near the high power red and blue curves of Fig. \ref{fig:dark_bright_soliton}(a). The SF approaches the lower power plateau displaying a decaying oscillatory trajectory, whilst the approach to the higher power plateau is smooth. It is the interactions of these oscillations on the low power plateau which introduce a locking mechanism analogous to that described previously for vectorial dark-dark solitons. The corresponding bright solitons of Fig. \ref{fig:dark_bright_soliton}(d) are marked with an `x' in Fig. \ref{fig:dark_bright_soliton}(c). Here, the switching fronts display a decaying oscillatory trajectory at the connection with the higher power plateau near the blue curve, and a smooth trajectory on approach to the lower power plateau near the red curve of Fig. \ref{fig:dark_bright_soliton}(a). The bright solitons `mirror' the dark solitons in profile (apart from a magnitude rescaling), exhibiting an identical size and number of peaks, due to local cross-phase modulation. The three VDBSs of Fig. \ref{fig:dark_bright_soliton}(d) display different locking distances (soliton widths) corresponding to the distinct cycles of the oscillatory tails, with three (red solid), two (black dashed), and one (blue dash-dot) peaks (troughs) in the bright (dark) soliton. Typical of systems of counterpropagating field components, VDBSs do not stem from the HSSs of this system, but instead stem from plateaus whose existence and stability is dependent on the size and number of VDBS. It can be seen that the powers of the plateaus of the three bistable VDBSs are different. This is attributed to the difference in nonlocal coupling of fields with different soliton widths.

VDBS solutions are distributed along snaking curves whose branches of stable VDBS solutions are shown as pink curves in Fig. \ref{fig:dark_bright_soliton}(b)-(c), each branch corresponding to a distinct VDBS size. Due to the presence of nonlocal terms, we expect the collapse snaking of the VDBS solutions to be tilted in parameter space \cite{Firth07,campbell2023dark,campbell2022counterpropagating}, although this effect is small due to the counteracting contributions of the dark and bright solitons to the averaging terms. As the VDBS solutions of Eqs. (\ref{eq:FP}) form around the symmetry broken HSS horn, they are found at much larger input powers than the symmetric dark-dark vectorial solitons discussed in the previous sections.

In Fig. \ref{fig:dark_bright_soliton}(e) the dual frequency combs corresponding to the VDBS of the red curves in Fig. \ref{fig:dark_bright_soliton}(d) are presented. VDBSs produce dual combs \cite{zhang2022dark,suh2016microresonator, dutt2018chip,coddington2016dual} at the output, where the blue and red combs corresponds to the dark and bright solitons of the two orthogonal polarizations, respectively. These two combs have identical spectral line spacing, corresponding to the cavity round trip time, but are symmetry broken in the spectral line power and in the spectral envelope. The difference in the power of the spectral lines is due to the much higher circulating power in the polarization field component supporting the dark soliton when compared to that one supporting the bright soliton. Modulations in the spectral envelope correspond to the oscillations on the peak of the bright soliton and on the trough of the dark soliton. These oscillations `mirror' each other in the field envelope producing displaced modulations in the two spectral envelopes.

\subsection{Nonlocal coupling of oscillating vectorial dark-bright solitons}

\begin{figure}
    \centering
    \includegraphics[width=1\linewidth]{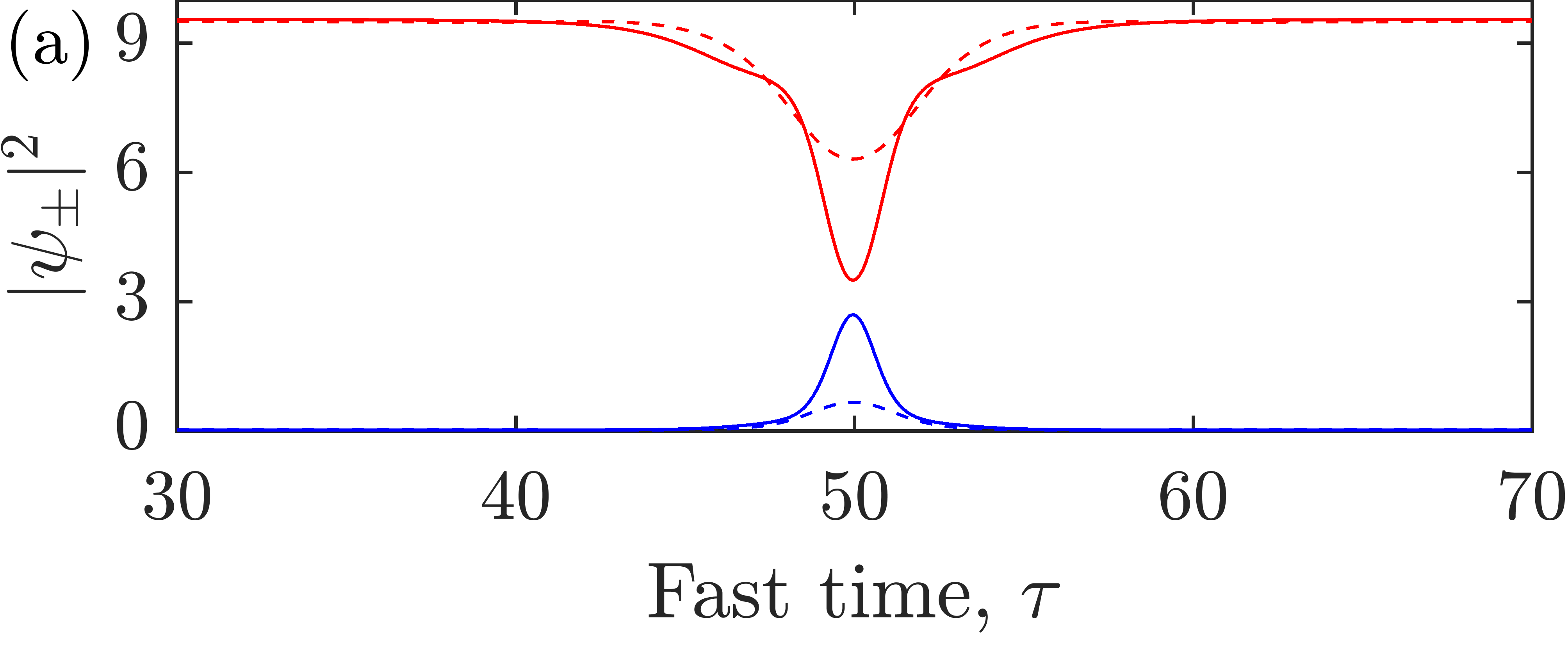}
    \includegraphics[width=1\linewidth]{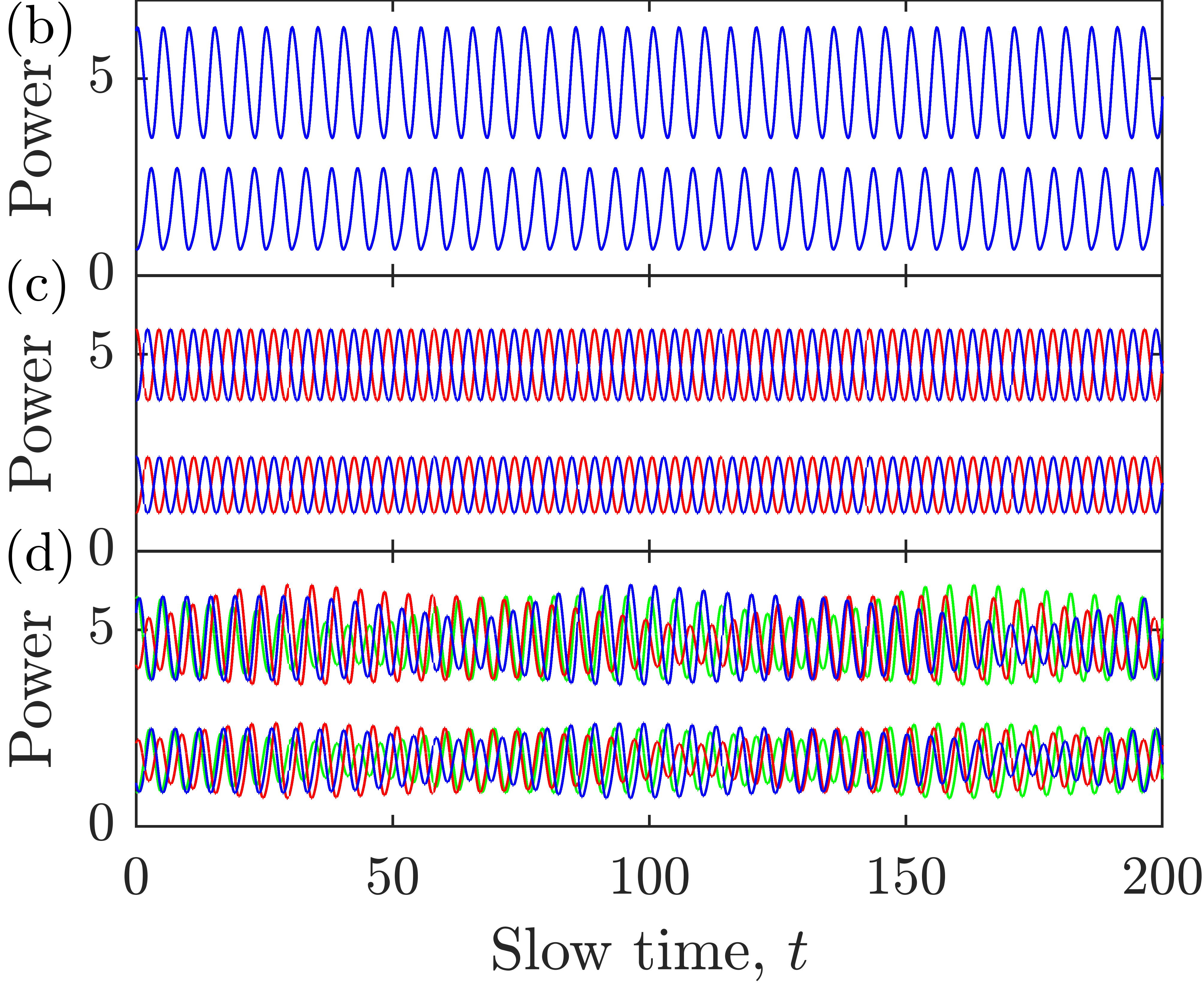}
    \includegraphics[width=1\linewidth]{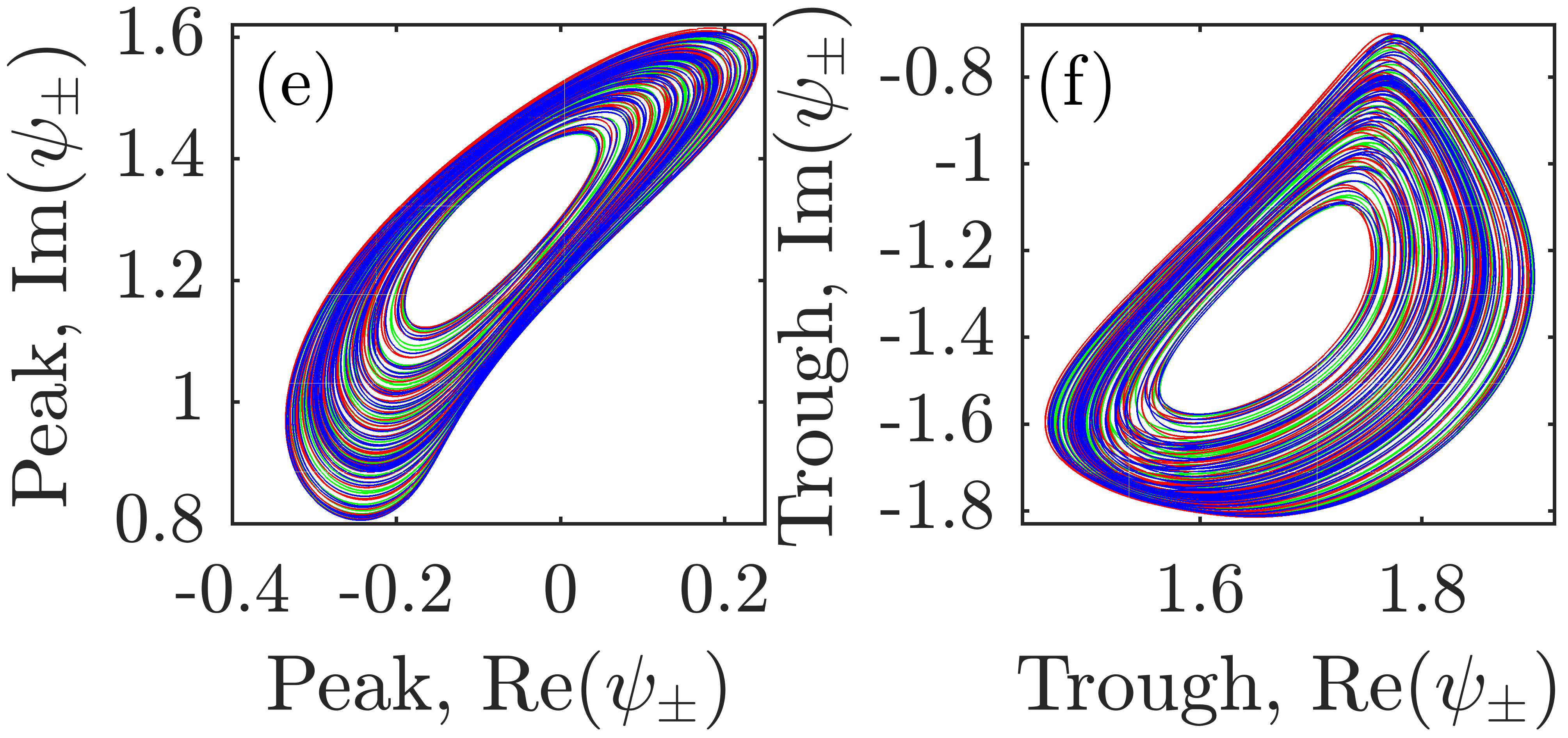}
    \caption[Anti-synchronization of dark-bright vectorial solitons]{Dynamical solitons for $S = 3.2$ and $\theta = 18.6$. (a) Power profile of a single oscillating VDBS at two different instants in its cycle, solid line and dashed line. (b) Time trace of the peak (trough) of the bright (dark) soliton of a single VDBS for $\tau_\text{R}=100$. (c) Time trace of the bright and dark solitons of two VDBS distinguished by different colored curves, for $\tau_\text{R}=200$. Nonlocal interactions between VDBS results in out of phase oscillations. (d) Time trace of the bright and dark solitons of three VDBS distinguished by different colored curves, for $\tau_\text{R}=300$. (e) Trajectory of bright soliton peaks in the Argand plane corresponding to (d). (f) Trajectory of dark soliton troughs in the Argand plane corresponding to (d). In both cases, the red, blue and green VDBS follow nontrivial nonlinear trajectories.}
    \label{fig:DBSPtrace_coupling}
\end{figure}

\begin{figure*}
    \centering\includegraphics[width=1\linewidth]{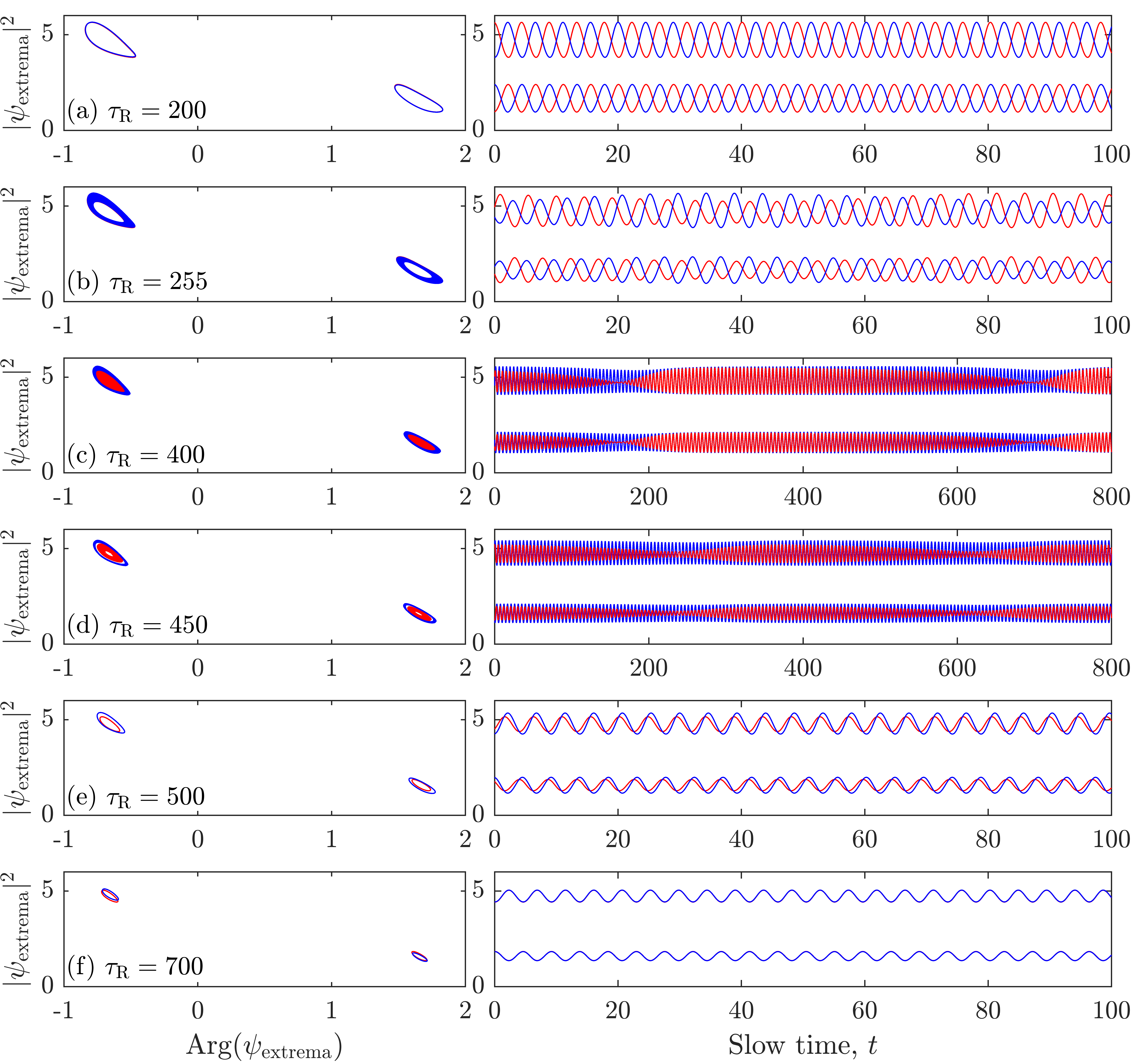}
    \caption[]{A pair of dynamical VDBSs for $S = 3.2$ and $\theta = 18.6$, and for different cavity round trip times, $\tau_\text{R}$. Left column: Trajectory of the minimum and maximum power of dark and bright solitons, respectively, for two oscillating VDBSs. Right column: The corresponding time trace of the extrema power over the slow time, where in each plot the two VDBS are distinguished by different colored curves. Nonlocal interaction strength between VDBS is weighted by the round trip time, such that, by increasing $\tau_\text{R}$ VDBS move from regimes of anti-phase (a) to in-phase (f) limit cycle oscillations, with nontrivial limit torus oscillations for intermediate values of $\tau_\text{R}$, (b)-(d).}
    \label{fig:twoVDBSdynamics}
\end{figure*}

When slowly changing the control parameters, the VDBS are found to undergo a Hopf bifurcation, resulting in breathing dynamics of the dark and bright solitons. The power profile of a single oscillating VDBS is shown in Fig. \ref{fig:DBSPtrace_coupling}(a) for $S=3.2$, $\theta=18.6$ and $\tau_\text{R} = 100$ at two instants (solid and dashed lines) along its dynamical cycle, corresponding to the maximum and minimum powers of the peak (trough) of the bright (dark) soliton. As the dark and bright solitons move along their limit cycle, the dynamics induce a small change in the nonlocal (averaged) terms of Eqs. (\ref{eq:FP}). As a consequence, the temporal dynamics of this system are not confined to the regions of the VDBS but extend the oscillation over the background plateaus far from the VDBS. Oscillations of the background power have identical periods but significantly smaller amplitudes than those of the VDBS peak. In Fig. \ref{fig:DBSPtrace_coupling}(b) we trace the maximum power of the bright soliton and and minimum power of the dark soliton over the slow time. One can see that the dark and bright solitons are perfectly synchronized in their dynamics. Fig. \ref{fig:DBSPtrace_coupling}(c) and (d) show these maxima and minima for a cavity containing two (c) and three (d) VDBSs oscillating simultaneously within the cavity of round trip times $\tau_\text{R} = 200$ and $\tau_\text{R} = 300$, respectively. Note that we increase the cavity round trip time for cavities containing a larger number of simultaneous solitons to preserve the value of the nonlocal terms of Eqs. (\ref{eq:FP}) at stationary state, so as to elucidate the effects of the nonlocal coupling on the soliton dynamics without affecting the existence and stability of the VDBS stationary solutions when we increase the soliton number. 

In Fig. \ref{fig:DBSPtrace_coupling}(c), we plot the maximum and minimum powers of the bright and dark solitons for the case with two VDBSs with red curves for the `right' VDBS and blue curves for the `left' VDBS. The two VDBSs interact through the nonlocal terms resulting in stable anti-phase dynamics, where the two bright solitons (and the two dark solitons) oscillate exactly $\pi$ out of phase. We have previously shown that the dynamics of $N$ well separated dark solitons in a FP resonator with round trip time $\tau_\text{R}$ and a single field component synchronized towards the dynamics of a single soliton in a cavity of round trip time $\tau_\text{R}/N$ \cite{campbell2023dark}. From Fig. \ref{fig:DBSPtrace_coupling}(c) it is clear that this is not the case for VDBSs of the two component FP resonator. This fact is further demonstrated when considering three VDBS well separated within a FP cavity with round trip time $\tau_\text{R}=300$. We plot in Fig. \ref{fig:DBSPtrace_coupling}(d) the time traces of the extrema of three VDBSs, distinguished by different colors. We see that the asymptotic dynamics of the three oscillating VDBS is nontrivial. When plotting the trajectory of the maxima of each bright soliton and the minima of each dark soliton in the Argand plane for the three VDBSs case, see Figs. \ref{fig:DBSPtrace_coupling}(e) and (f), we can see that the dynamics is quasiperiodic leading to a limit torus for all three bright and dark VDBSs.

\begin{figure}
    \centering\includegraphics[width=1\linewidth]{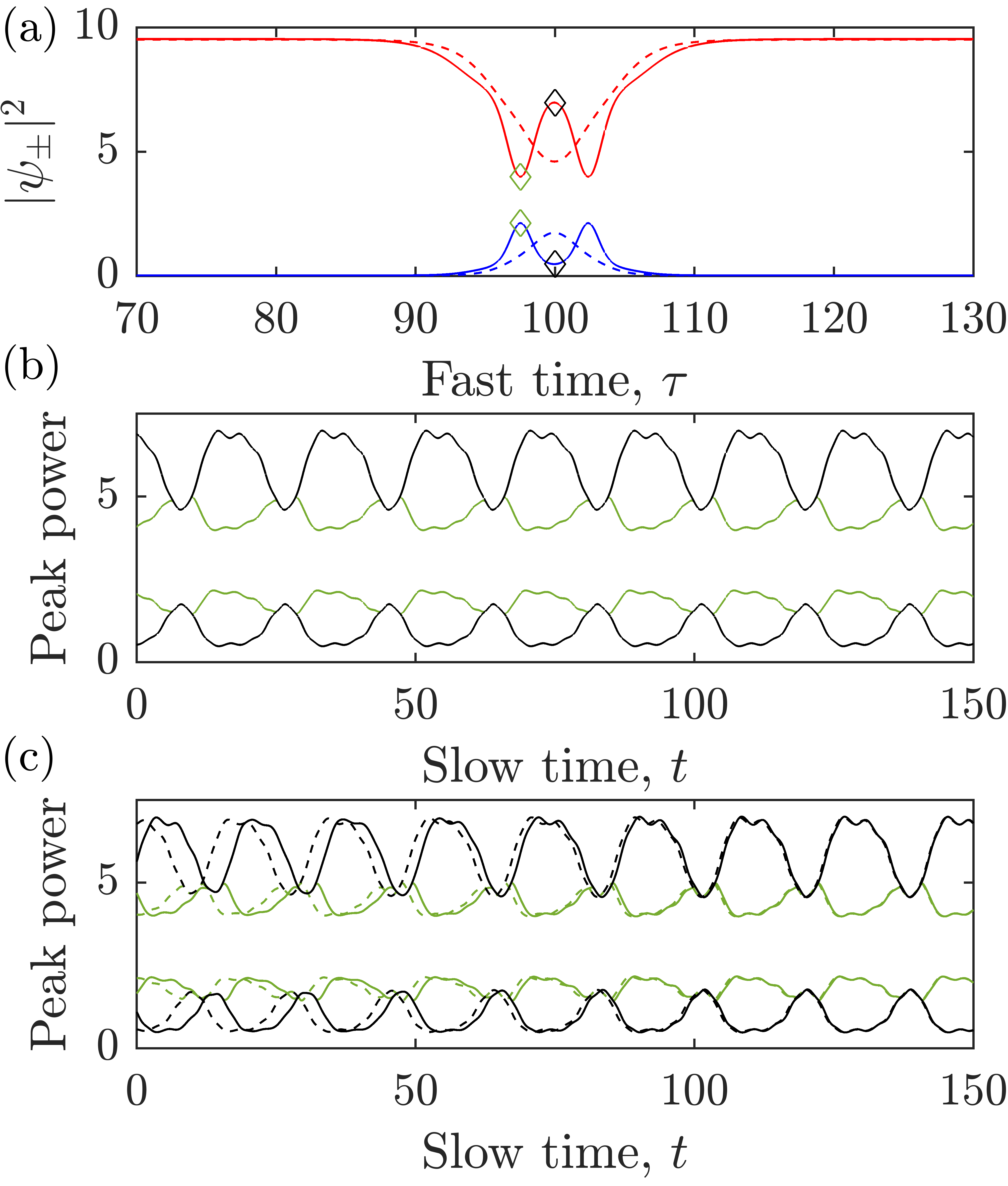}
    \caption[Synchronization of multi-peek dark-bright vectorial solitons]{Dynamical VDBS for $S = 3.2$ and $\theta = 18.6$. (a) Power profile at two instants in slow time of the VDBS oscillating between one and two peaks. (b) Time traces of the off centre trough/peak (green line) and central trough/peak (black line) of a single VDBS for $\tau_\text{R} = 200$. (b) Time traces of the off center trough/peak (red) and central trough/peak (blue) of two VDBS for $\tau_\text{R} = 400$. The solid (dashed) lines correspond to the first (second) VDBS. The two VDBS oscillate out of phase initially, but synchronize over time due to nonlocal coupling. The tracking points of the trace are indicated with diamonds in (a).}
    \label{fig:DBSPtrace_couplingtwopeak}
\end{figure}

If instead of considering $N$ solitons in cavities with round trip $N\tau_\text{R}$ as done above, we consider the oscillations of two VDBSs in cavities with increasing round trip times $\tau_\text{R}$, we can investigate the contribution of the integral (nonlocal) terms to their dynamics. In general, the strength of the interaction between nonlocally coupled VDBSs depends on the density of VDBSs over the round trip time, and as such, the interaction strength of two VDBSs is weighted by the size of $\tau_\text{R}$ with weaker interaction in the limit of large $\tau_\text{R}$. In Fig. \ref{fig:twoVDBSdynamics}, we plot the trajectories of the extrema of two VDBSs for different values of the round trip time, where the left column shows the trajectory of the peak (trough) of the bright (dark) solitons in the Argand plane, respectively, while the right column shows the trace of the peak (trough) power of the bright (dark) solitons over the slow time when changing $\tau_\text{R}$. In Fig. \ref{fig:twoVDBSdynamics}(a), VDBSs display anti-phase and overlapping limit cycle oscillations (see also Fig. \ref{fig:DBSPtrace_coupling}(c)). When increasing the round trip time to $\tau_\text{R}=255$, limit cycle oscillation give way to stable limit torus oscillations, seen in \ref{fig:twoVDBSdynamics}(b), where VDBSs oscillate along overlapping trajectories with small periodic variations in their phase and amplitude over the slow time. This overlap of trajectories eventually gives way to VDBSs which evolve along distinct trajectories, as can be seen in Fig. \ref{fig:twoVDBSdynamics}(c) for the round trip time $\tau_\text{R} = 400$. Here, one of the VDBSs displays dominant dynamics (blue curve), oscillating with near steady amplitude and phase, while the second VDBS exhibits oscillations that grow and decay in amplitude (red curve). We see that the amplitude of the submissive VDBS is dependent on the mismatch of oscillation phase between the two VDBSs, where the red VDBS oscillates with large amplitude when in phase with the blue VDBS. As the VDBSs evolve, their oscillations move progressively out of phase, such that, the amplitude of oscillation approaches zero, at which point the system rapidly moves back to equal amplitude and in phase oscillations. These slow time dynamics then repeat. 
For the sightly larger round trip time $\tau_\text{R} = 450$, in Fig. \ref{fig:twoVDBSdynamics}(d), the VSBSs move along distinct non-overlapping limit tori. Here, the submissive VDBS shows repeated times of growth and decay in amplitude as the VDBSs move in and out of phase periodically. For even larger cavities round trip times, the dynamics of the VDBSs return to stable limit cycle oscillation. In such regimes, the two VDBS oscillate with constant amplitudes and identical frequency, with a constant phase offset as shown in Fig. \ref{fig:twoVDBSdynamics}(e) for $\tau_\text{R}= 500$. In the limit of large $\tau_\text{R}$, these limit cycle oscillations approach full symmetry by displaying identical limit cycle trajectories, see Fig. \ref{fig:twoVDBSdynamics}(f). This suggests that VDBSs favor synchronization in the limit of a large cavity round trip time, anti-phase dynamics for a short cavity round trip time, and nontrivial limit torus dynamics for intermediary values. Some of these features are reminiscent of self-switching behaviours seen in the homogeneous case of the SSB of counter-propagating fields in a ring resonator \cite{Woodley21}.

The distinct dynamical regimes found when increasing $\tau_\text{R}$ can be explained by considering the change in the relative sizes of the real valued nonlocal terms (Eq. (\ref{eq:inttermreal})) and the complex valued nonlocal term (Eq. (\ref{eq:inttermcomp})) over slow time. As demonstrated for the single field FP model \cite{campbell2023dark}, real valued nonlocal terms are associated with in-phase dynamics, regardless of cavity size. As such, we may attribute anti-phase dynamics to the unique feature of Eqs. (\ref{eq:FP}), namely the complex valued nonlocal term. Oscillations of the dark and bright solitons give opposite contributions to the average power and hence result in relatively small changes in the real valued integral terms. For small cavity sizes, the dynamics are then dominated by the complex valued nonlocal term (\ref{eq:inttermcomp}), but as the cavity size increases, the change in (\ref{eq:inttermcomp}) decreases until the contribution of the real valued integrals become dominant and in-phase dynamics are attained.

This can be further demonstrated by considering the dynamics of a VDBS of larger width. In Fig. \ref{fig:DBSPtrace_couplingtwopeak}(a), we present the power profile of a VDBS at two points in its dynamics cycle, as it oscillates between two peaks/troughs and one peak/trough. This dynamical VDBS is found for identical parameter values of the anti-phase VDBSs of Fig. \ref{fig:DBSPtrace_coupling}(c); they are hence bistable. The time trace of the central peak/trough and the two side peaks/troughs are plotted in Fig. \ref{fig:DBSPtrace_couplingtwopeak}(b) as black and green curves, respectively, and show the stable limit cycle oscillation of the VDBS. If we introduce an additional VDBS into the cavity for $\tau_\text{R} = 400$, we find that the VDBSs move towards synchronous dynamics, as can be seen in Fig. \ref{fig:DBSPtrace_couplingtwopeak}(c). Comparing the dynamics of Figs. \ref{fig:DBSPtrace_coupling} and \ref{fig:DBSPtrace_couplingtwopeak}, we see that the particular profile of the oscillating VDBSs contributes to whether the long range interactions result in in-phase or anti-phase dynamics. In particular, we note that the dynamics of Fig. \ref{fig:DBSPtrace_couplingtwopeak}(c) induces a significantly larger change in the real valued integral terms, when compared with the single peak VDBSs. This is due to more complicated oscillations and leads to larger variations in the soliton width (pulse duration), associated with moving between one and two peaks. As such, in-phase dynamics become dominant for VDBSs with this broader profile for much smaller values of $\tau_\text{R}$.


\section{Vectorial cavity soliton distribution in parameter space}
\begin{figure}
    \centering\includegraphics[width=1\linewidth]{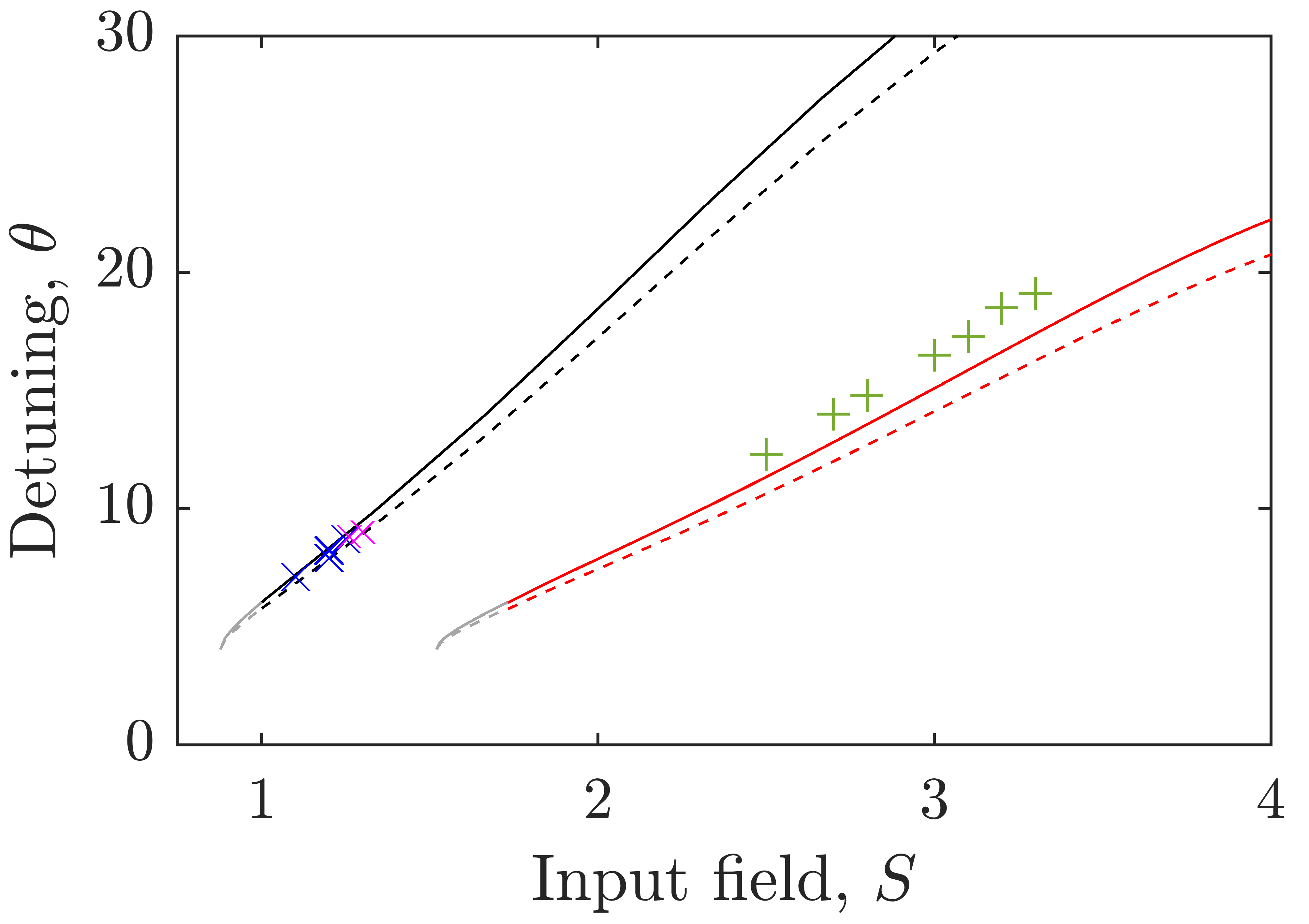}
    \caption[Dark and dark-bright vectorial soliton distribution in parameter space]{Dark-dark and dark-bright vectorial soliton distribution in parameter space. Example VDS are plotted as x's and VDBS as +'s. Semi-analytical Maxwell points solitons at zero dispersion are shown at symmetry for $\Delta \rightarrow 0$ (black line) and $\Delta = 0.1$ (black dashed line). By using the single field Eq. (\ref{eq:brightfieldzero}) for VDBS, semi-analytical Maxwell point solitons at zero dispersion are shown for $\Delta \rightarrow 0$ (red line) and $\Delta = 0.1$ (red dashed line).}
    \label{fig:solutionmap}
\end{figure}

In a way analogous to the results of \cite{campbell2022counterpropagating,campbell2023dark}, the soliton solutions of Eqs. (\ref{eq:FP}) are related to the presence of Maxwell points in parameter space. For the LLE, the Maxwell point corresponds to a single value of the detuning corresponding to SFs with zero velocity. This results in a multitude of stationary solutions composed of SF with arbitrary separations $\Delta$ \cite{parra2016dark,campbell2022counterpropagating}. Maxwell points in Eqs. (\ref{eq:FP}) allow us to predict the location of VDS and VDBS. Due to the nonlocal terms in the FP resonator model, Maxwell point solutions are tilted in parameter space \cite{Firth07,campbell2022counterpropagating,campbell2023dark}, such that, for each value of the detuning, there exists a single separation $\Delta$ of two SFs where the SFs have zero velocity. In order to predict the formation of VDSs, we plot in Fig. \ref{fig:solutionmap} the Maxwell points for separations $\Delta \rightarrow 0$ (black solid curve) and $\Delta = 0.1$ (black broken curve). These two curves provide approximate estimates of where to find soliton solutions in the full equations since they are obtained at zero dispersion and with discontinuous walls. Examples of stable symmetric VDS solutions are shown in Fig. \ref{fig:solutionmap} with x's and show good agreement with the predictions through the Maxwell points.

We also note that VDBSs are highly symmetry-broken solutions where one field is largely suppressed. This is most prominent in the background plateaus of the bright solitons which approaches almost zero power as can be seen in Fig. \ref{fig:dark_bright_soliton}(d). Due to the large disparity in power of the fields containing dark and bright solitons, we can consider the bright soliton background field to act perturbatively on the field supporting the dark soliton. If we assume the background field of the bright soliton to be close to zero $\psi_-\rightarrow0$ at all points in the cavity, we can approximate the equation governing the field supporting the dark soliton as 
\begin{equation}
    \partial_t \psi_+ = S - (1+i\theta)\psi_+ + i\frac{2}{3}\{|\psi_+|^2 + 2\langle|\psi_+|^2\rangle\}\psi_+ - i\partial_\tau^2\psi_+ \, .\label{eq:brightfieldzero}
\end{equation}
This equations is identical to the single field FP model under renormalization. Eq. (\ref{eq:brightfieldzero}) is only an approximation to the field $\psi_+$ of Eqs. (\ref{eq:FP}), but if we plot the corresponding Maxwell point in Fig. \ref{fig:solutionmap} (red curve), we are able to predict the location in parameter space of the VDBSs. Examples of VDBSs solutions obtained in simulation are indicated with +'s, and can be seen to follow the predicted curve from Eq. (\ref{eq:brightfieldzero}).


\section{Conclusion}
We investigated the propagation of light composed of counter-rotating circular polarization components within a FP cavity filled with a Kerr nonlinear medium. Following from \cite{hill2024symmetry}, this system is governed by two coupled integro-partial differential equations, describing intracavity fields of orthogonal polarization, which interact through self- and cross-phase modulation terms. At difference to \cite{hill2024symmetry}, here we investigated the regime of normal group velocity dispersion. Comb generation across different dispersion regimes can increase the spectral bandwidth of the frequency combs for practical applications \cite{Pal23}. 

Linear stability analysis of our model yielded a bifurcation structure of competing SSB bifurcations. We presented the SSB bifurcations of the homogeneous stationary states, where in particular, we characterized a codimension-2 bifurcation point of the symmetric HSSs unique to the regime of normal group velocity dispersion. The codimension-2 bifurcation represents a point in parameter space in which a reverse pitchfork bifurcation of the symmetry broken HSSs and a forward pitchfork bifurcation of a Turing instability collide, with the real part of the corresponding eigenvalues being simultaneously zero. SSB of the HSS is associated with the reverse pitchfork bifurcation, and has been previously demonstrated experimentally when neglecting dispersion \cite{moroney2022Kerr}. Here, we characterized the forward pitchfork bifurcation, innate to the normal dispersion regime, leading to Turing patterns composed of fields with alternating orthogonal polarization. This bifurcation structure extends to the homogeneous background of a symmetric VDS, resulting in a multitude of symmetry broken VDS solutions. Of particular interest, is a SSB bifurcation of the VDS resulting in the formation of Turing patterns on the homogeneous background from which the soliton hangs.

When considering multiple simultaneous VDSs along the cavity round trip, SSB Turing patterns are found to form in the intervals between adjacent VDSs. As the Turing patterns grow in amplitude, VDSs are `pushed' along the resonator, increasing their separation until equidistant equilibrium distances are reached. In other words, the SSB of VDSs induce long range repulsive interactions between adjacent VDSs, mediated by the Turing patterns. The motion of the VDSs represents a spontaneous self-organization phenomenon, which results in the formation of a regularly spaced soliton crystal. The `self-crystallization' of a RSC was previously demonstrated for VDSs of the ring resonator in \cite{campbell2024frequency}. In comparison, the Turing patterns of the FP resonator present with a higher wavenumber and longer amplitude saturation distance resulting in an increased range of the repulsive interaction between VDSs in the FP cavity. Hence, the FP resonator is more congruent to the formation for RSCs and more robust to changes in soliton number. RSCs originate spontaneously from a random distribution of VDS, without the need of a perturbation \cite{cole2017soliton,lu2021synthesized,wang2018robust,karpov2019dynamics,lin2022mode}, offering new degrees of control, and a simple implementation, relevant for applications in RSC generation \cite{federici2010review,riemensberger2020massively,xu2018advanced,hu2020reconfigurable}. The generalization of this process to the FP resonator increases access to this phenomenon for practical application. RSCs produce a frequency comb with smooth spectral profile and an increased line spacing when compared to a random distribution of cavity solitons. As such, a RSC may be used to emulate smaller cavity sizes while avoiding the associated experimental limitations. Note that conversion efficiency of dark cavity solitons to frequency combs can be much better compared to bright solitons, which is important for reducing the power consumption of chip integrated comb generators \cite{Jang21,Helgason23,Yang24}.

Finally, we characterized the formation of VDBSs in normal dispersion regimes. These solutions form in regimes of highly symmetry broken HSSs, in which, the circulating power in one polarization component is significantly larger than the power of the other. With suitable perturbations, a dark soliton forms in the high power polarization component, which is coupled to a bright soliton that forms in the low power component. Previous methods for the generation of VDBSs make use of bichromatic driving of a ring resonator \cite{zhang2022dark}, such that the two frequency components may operate in distinct regimes of group velocity dispersion, with one laser operating in the anomalous dispersion regime, generating a bright soliton, while the other operates in the normal dispersion regime, supporting a dark soliton though cross-phase modulation. In our system, instead, bright and dark solitons are formed in regimes of identical normal dispersion, and are composed of switching fronts which connect two high power plateaus and two low power plateaus. SFs lock to form solitons due to the interaction of oscillatory tails which appear on the approach to the plateaus of intermediary power. VDBS of the FP resonator can undergo Hopf bifurcations when scanning the detuning, resulting in breathing dynamics. We found that well separated VDBSs, located such that they do not interact via the local dynamics at the tails, experience oscillation-phase dependent interactions through the nonlocal coupling. A pair of well separated dynamical VDBSs can oscillate exactly out of phase with overlapping limit cycle trajectories. This is unlike the in-phase dynamics of single component dark cavity solitons of the FP model in \cite{campbell2023dark}. Conversely, in the limit of a large cavity round trip time, the VDBS dynamics found to approach synchronization. In general, short cavities display anti-phase dynamics, long cavities display in-phase dynamics, and intermediate cavities display nontrivial dynamics. These regimes are a consequence of competition between real valued and complex valued nonlocal coupling terms. We also note the possibility of VDBS solutions that are analogous to those of the ring resonator system, where due to the absence of nonlocal coupling, well separated VDBS do not interact.


\vfill

We acknowledge support from the EPSRC DTA Grant No. EP/T517938/1. P.D. acknowledges support by the European Union's H2020 ERC Grant ``CounterLight'' 756966 and the Max Planck Society. LH acknowledges support from the SALTO scheme of the Max-Planck-Gesellschaft (MPG) and the CNRS.

\vfill

\bibliographystyle{unsrt}

\appendix
\begin{widetext}

\newpage\section{Linear stability analysis of the homogeneous stationary state.}\label{app:stabHSS}
To investigate the linear stability of the HSSs, we perform a model expansion of the field envelope
\begin{eqnarray}
    \psi_\pm(\tau,t) = \sum_{\mu=-\infty}^\infty f_\mu^\pm(t)e^{i\alpha_\mu\tau},
\end{eqnarray}
and insert it into Eqs. (\ref{eq:FP}) to yield the modal equations
\begin{align}
    \partial_t f^{\pm}_\mu
    =~&S\delta_{\mu,0} - (1+i\theta){f}^{\pm}_\mu + i\alpha_\mu^2{f}^{\pm}_\mu\nonumber\\
    & +\frac{2i}{3}\sum_{\mu',\mu'',\mu'''}\delta_{\mu,\mu'+\mu''-\mu'''}{f}^\pm_{\mu'}{f}^\pm_{\mu''}({f}^\pm_{\mu'''})^* + \frac{4i}{3}{f}^\pm_{\mu}\sum_{\mu'}({f}^\pm_{\mu'})^*{f}^\pm_{\mu'} \nonumber\\
    & +\frac{4i}{3}\sum_{\mu',\mu'',\mu'''}\delta_{\mu,\mu'+\mu''-\mu'''}{f}^\pm_{\mu'}{f}^\mp_{\mu''}({f}^\mp_{\mu'''})^* + \frac{4i}{3}{f}^\pm_{\mu}\sum_{\mu'}{f}^\mp_{\mu'}({f}^\mp_{\mu'})^*\nonumber\\
    & +\frac{4i}{3}{f}_{\mu}^\mp\sum_{\mu'}({f}_{\mu'}^\mp)^*{f}_{\mu'}^\pm,\label{eq:modalFP}
\end{align}
where $f_\mu(t)$ is the modal amplitude with cavity mode number $\mu$ and fast time wavenumber $\alpha_\mu^2 = (2\pi\mu/\tau_\text{R})^2$. We then perform a linearization of the modal amplitudes about the HSSs of Eqs. (\ref{eq:FP}), which correspond to the modal coefficients 
\begin{equation}\label{eq:modalHSSFPtwo}
    f_{\mu,s}^\pm = \psi_s^\pm\delta_{\mu,0}
\end{equation}
where $\delta_{\mu,0}$ is the Kronecker delta. We introduce a linear perturbation to the the counter-rotating polarization components of the form
\begin{equation}
    f_\mu^\pm = \psi_s^\pm \delta_{\mu,0} + \delta f_\mu^\pm,
\end{equation}
which is inserted into Eqs. (\ref{eq:modalFP}), yielding the linearized set of equations
\begin{align}
    \partial_t \delta f^{\pm}_\mu
    &= - (1+i\theta)\delta{f}^{\pm}_\mu + i\alpha^2_\mu\delta{f}^{\pm}_\mu\label{eq:perturb_equation}\\
    &~ +\frac{2i}{3}\{(4|{\psi}^\pm_{s}|^2+4|{\psi}^\mp_{s}|^2)\delta {f}^\pm_{\mu} + 4{\psi}^\pm_{s}({\psi}^\mp_{s})^*\delta f^\mp_{\mu} + ({\psi}^\pm_{s})^2(\delta f^\pm_{-\mu})^* + 2{\psi}^\pm_{s}{\psi}^\mp_{s}(\delta f^\mp_{-\mu})^*\} \nonumber\\
    &~  + \frac{4i}{3}\delta_{\mu,0}\{(|{\psi}^\pm_{s}|^2 + |{\psi}^\mp_{s}|^2)\delta{f}^\pm_{0} + {\psi}^\pm_{s}({\psi}^\mp_{s})^*\delta{f}^\mp_{0} + ({\psi}^\pm_{s})^2(\delta{f}^\pm_{0})^* + 2{\psi}^\pm_{s}{\psi}^\mp_{s}(\delta{f}^\mp_{0})^*\}.\nonumber
\end{align}
Without loss of generality, we may assume that the homogeneous stationary solutions are real. Hence we express the real and imaginary components of the perturbations as 
\begin{equation}\label{eq:FP2poljacpert}
    \begin{pmatrix}
    \partial_t\operatorname{Re}(\delta f_\mu^+)\\
    \partial_t\operatorname{Im}(\delta f_\mu^+)\\
    \partial_t\operatorname{Re}(\delta f_\mu^-)\\
    \partial_t\operatorname{Im}(\delta f_\mu^-)
    \end{pmatrix}
    =
    \begin{pmatrix}
    -1   &   -A_+   &   0   &   -\frac{1}{6}(1-\delta_{\mu,0})C\\
    B_+   &   -1   &   \frac{1}{2}C   &   0\\
    0   &   -\frac{1}{6}(1-\delta_{\mu,0})C   &   -1 &   -A_-\\
    \frac{1}{2}C   &    0   &   B_-   &   -1
    \end{pmatrix}
    \begin{pmatrix}
    \operatorname{Re}(\delta f_\mu^+)\\
    \operatorname{Im}(\delta f_\mu^+)\\
    \operatorname{Re}(\delta f_\mu^-)\\
    \operatorname{Im}(\delta f_\mu^-)
    \end{pmatrix},
\end{equation}
where
\begin{align}
    A_\pm &= -\theta + \alpha_\mu^2 + 2\psi_\pm^2 + \frac{4}{3}(2+\delta_{\mu,0})\psi_\mp^2,\\
    B_\pm &= -\theta + \alpha_\mu^2 + \frac{2}{3}(5+4\delta_{\mu,0})\psi_\pm^2 + \frac{4}{3}(2+\delta_{\mu,0})\psi_\mp^2,\\
    C^2 &= 64(1+3\delta_{\mu,0})\psi_+^2\psi_-^2.
\end{align}
The Jacobian matrix of Eq. (\ref{eq:FP2poljacpert}) yields the characteristic equation,
\begin{align}
    0 = (\lambda + 1)^4 &+ \bigg[A_+B_+ + A_-B_- + \frac{1}{6}(1-\delta_{\mu,0})C^2\bigg](\lambda + 1)^2\nonumber\\ 
    &+ A_+A_-B_+B_- - \frac{1}{4}A_+A_-C^2 - \frac{1}{36}(1-\delta_{\mu,0})C^2\bigg[B_+B_- + 
    \frac{1}{4}C^2\bigg],
\end{align}
to obtain the eigenvalues \cite{hill2024symmetry}
\begin{align}
    \lambda(\alpha_\mu) = -1 \pm \frac{\sqrt{-A_+B_+ - A_-B_- - \frac{1}{6}(1-\delta_{\mu,0})C^2\pm Q}}{\sqrt{2}},
\end{align}
with
\begin{align}
    Q = \sqrt{(A_+B_+ - A_-B_-)^2 + A_+A_-C^2 + (1-\delta_{\mu,0})\frac{C^2}{9}(3A_+B_+ + B_+B_- + 3A_-B_-)}.
\end{align}
These eigenvalues are separated into regimes of $\mu = 0$ and $\mu\neq0$ due to the presence of the Kronecker delta. The stability of the regimes where these eigenvalues apply, can be heuristically understood by considering the effect of the perturbation of the integral term,
\[
    \langle (\psi_s^a + \epsilon^a)(\psi_s^b + \epsilon^b)^*\rangle \sim 
\begin{cases}
    (\psi_s^a+\epsilon^a)(\psi_s^b+\epsilon^b)^*,& \text{if } \epsilon \propto \exp[\lambda(\alpha_0=0) t], ~(\mu = 0),\\
    \psi_s^a(\psi_s^b)^*,& \text{if } \epsilon \propto \exp[\lambda(\alpha_\mu) t + i\alpha_\mu \tau],~(\mu \neq 0).
\end{cases}
\]
where $a,b$ indicate the polarizations $\pm$. Should the perturbation take a sinusoidal form over the cavity coordinate, it will not survive the integral terms. The consequences of these eigenvalues are discussed in Section \ref{sec:FPSSB_HSS} of the main text.

\section{Linear stability analysis of step functions to pattern formation}\label{app:stabplat}
We can investigate the formation of Turing patterns on symmetric VDS solutions by approximating the VDSs as two plateaus occupying distinct domains of fast time connected by step functions $\psi^\text{s} = \psi^\text{s}_{u} + \psi^\text{s}_{l}$. We introduce a linear perturbation to each plateau of the form 
\begin{align}
    \psi_{\pm,u}(\tau,t) &= \psi_u^\text{s} + \epsilon_{\pm,u} e^{i\alpha_u \tau+\Omega_u t}\label{eq:FP2polPlatPertuapp}\\
    \psi_{\pm,l}(\tau,t) &= \psi_l^\text{s} + \epsilon_{\pm,l} e^{i\alpha_l \tau+\Omega_l t}\label{eq:FP2polPlatPertlapp}
\end{align}
where $\psi_{u,l}^s$ are the plateau solutions of higher $u$ and lower $l$ power, $k_{u,l}$ and $\Omega_{u,l}$ are the wavenumbers and growth rates of the perturbation on the respective plateaus, and $\epsilon_{\pm,u}, \epsilon_{\pm,l}<<1$. Inspired by the linear stability analysis previously performed for the HSSs we assume that the perturbation does not survive the integral terms, such that perturbations (\ref{eq:FP2polPlatPertu}) and (\ref{eq:FP2polPlatPertl}) obey
\begin{eqnarray}
    \langle |\psi_+|^2\rangle = \langle |\psi_-|^2\rangle = \langle \psi_+\psi_-^*\rangle = \langle \psi_+^*\psi_-\rangle =  \langle |\psi^\text{s}|^2\rangle = \Delta (\psi_l^s)^2 + (1-\Delta)(\psi_u^s)^2.
\end{eqnarray}
This is akin to assuming that the wavenumbers $\alpha_{u,l}$ are periodic on their respective plateau. As the perturbations do not survive the nonlocal coupling terms, the respective perturbations on the upper and lower plateau are no longer coupled, so that, without loss of generality, we can consider the high and lower power plateaus to be real. Inserting the step function approximation into Eqs. (\ref{eq:FP}) and performing the linerizations (\ref{eq:FP2polPlatPertu}), (\ref{eq:FP2polPlatPertl}), we arrive at
\begin{equation}
    \begin{pmatrix}
        \partial_t u^{+}_{u}\\
        \partial_t v^{+}_{u}\\
        \partial_t u^{-}_{u}\\
        \partial_t v^{-}_{u}\\
        \partial_t u^{+}_{l}\\
        \partial_t v^{+}_{l}\\
        \partial_t u^{-}_{l}\\
        \partial_t v^{-}_{l}
    \end{pmatrix}
    =
    \begin{pmatrix}
    -1  &  A_u   &   0  & C_1  &  0   &   0   &   0   &   0\\
    -B_u  &  -1   &   -C_2 & 0  &  0   &   0   &   0   &   0\\
    0  &  C_1   &   -1  & A_u    &  0   &   0   &   0   &   0\\
    -C_2  &  0  &   -B_u & -1   &  0   &   0   &   0   &   0\\
    0   &   0   &   0   &   0   &   -1  &  A_l   &   0  & C_1\\
    0   &   0   &   0   &   0   &   -B_l  &  -1   &   -C_2 & 0 \\
    0   &   0   &   0   &   0   &   0  &  C_1   &   -1  & A_l\\
    0   &   0   &   0   &   0   &   -C_2  &  0  &   -B_l & -1
    \end{pmatrix}
    \begin{pmatrix}
        u^{+}_{u}\\
        v^{+}_{u}\\
        u^{-}_{u}\\
        v^{-}_{u}\\
        u^{+}_{l}\\
        v^{+}_{l}\\
        u^{-}_{l}\\
        v^{-}_{l}
    \end{pmatrix}
\end{equation}
in terms of the real and imaginary parts of the perturbations and we have $A_{u,l} = \theta-\alpha_{u,l}^2 - 2(\psi_{u,l}^s)^2 - 8/3\langle|\psi^s|^2\rangle$, $B_{u,l} = \theta'-\alpha_{u,l}^2 - 10/3(\psi_{u,l}^s)^2 - 8/3\langle|\psi^s|^2\rangle$, $C_1 = - 4/3\langle|\psi^s|^2\rangle$, $C_2 = C_1 - 8/3(\psi_{u,l}^s)^2$. ($\psi_{u,l}^s)^2$ are the powers of the high and lower power plateaus and $\langle|\psi^s|^2\rangle$ is the average of the intracavity field. This analysis yields eigenvalues
\begin{equation}    
    \Omega_{u,l}(k_{u,l}) = -1 \pm \sqrt{-A_{u,l}B_{u,l} -C_1C_2\pm (A_{u,l} C_2 + B_{u,l} C_1)}.\label{eq:pat_eigenvalues1app}
\end{equation}
These eigenvalues are discussed in Section \ref{sec:FPSSB_VDS} of the main text.
\end{widetext}

\end{document}